\documentstyle[preprint,aps,floats,epsf]{revtex}

\def\Re{{\cal R}\!e}
\def\Im{{\cal I}\!m}

\begin{document}
\tightenlines
%% the segment below is from prabib.sty, and has been altered so that a new
%% page is not started for the references (saves paper)
\catcode`@=11
\def\references{%
\ifpreprintsty
%\newpage
\bigskip\bigskip
\hbox to\hsize{\hss\large \refname\hss}%
\else
\vskip24pt
\hrule width\hsize\relax
\vskip 1.6cm
\fi
\list{\@biblabel{\arabic{enumiv}}}%
{\labelwidth\WidestRefLabelThusFar  \labelsep4pt %
\leftmargin\labelwidth %
\advance\leftmargin\labelsep %
\ifdim\baselinestretch pt>1 pt %
\parsep  4pt\relax %
\else %
\parsep  0pt\relax %
\fi
\itemsep\parsep %
\usecounter{enumiv}%
\let\p@enumiv\@empty
\def\theenumiv{\arabic{enumiv}}%
}%
\let\newblock\relax %
\sloppy\clubpenalty4000\widowpenalty4000
\sfcode`\.=1000\relax
\ifpreprintsty\else\small\fi
}
\catcode`@=12

\font\fortssbx=cmssbx10 scaled \magstep2
\hbox to \hsize{{\fortssbx University of Wisconsin - Madison}
\hfill\vtop{
\hbox{\bf MADPH-00-1199}
\hbox{\bf FERMILAB-Pub-00/318-T}
\hbox{\bf AMES-HET-00-12}
\hbox{\bf hep-ph/0012017}
\hbox{November 2000}}}

\vspace*{.25in}
\begin{center}
{\large\bf Exploring Neutrino Oscillations with Superbeams}\\[10mm]
V. Barger$^1$, S. Geer$^2$, R. Raja$^2$, and K. Whisnant$^3$\\[5mm]
\it
$^1$Department of Physics, University of Wisconsin,
Madison, WI 53706, USA\\
$^2$Fermi National Accelerator Laboratory, P.O. Box 500,
Batavia, IL 60510, USA\\
$^3$Department of Physics and Astronomy, Iowa State University,
Ames, IA 50011, USA

\end{center}
\thispagestyle{empty}

\begin{abstract}

\vspace*{-.35in}

\noindent
We consider the medium- and long-baseline oscillation physics
capabilities of intense muon-neutrino and muon-antineutrino beams
produced using future upgraded megawatt-scale high-energy proton
beams. In particular we consider the potential of these conventional
neutrino ``superbeams'' for observing $\nu_\mu\to\nu_e$ oscillations,
determining the hierarchy of neutrino mass eigenstates, and measuring
$CP$-violation in the lepton sector. The physics capabilities of
superbeams are explored as a function of the beam energy, baseline, and
the detector parameters (fiducial mass, background rates, and systematic
uncertainties on the backgrounds). The trade-offs between very large
detectors with poor background rejection and smaller detectors with
excellent background rejection are illustrated. We find that, with an
aggressive set of detector parameters, it may be possible to observe
$\nu_\mu\to\nu_e$ oscillations with a superbeam provided that the
amplitude parameter $\sin^2 2\theta_{13}$ is larger than a few $\times
10^{-3}$.  If $\sin^2 2\theta_{13}$ is of order $10^{-2}$ or larger,
then the neutrino mass hierarchy can be determined in long-baseline
experiments, and if in addition the large mixing angle MSW solution
describes the solar neutrino deficit then there is a small region of
parameter space within which maximal $CP$-violation in the lepton sector
would be observable (with a significance of a few standard deviations)
in a low-energy medium-baseline experiment. We illustrate our results by
explicitly considering massive water Cherenkov and liquid argon
detectors at superbeams with neutrino energies ranging from 1~GeV to
15~GeV, and baselines ranging from 295~km to 9300~km.  Finally, we
compare the oscillation physics prospects at superbeams with the
corresponding prospects at neutrino factories. The sensitivity at a
neutrino factory to $CP$ violation and the neutrino mass hierarchy
extends to values of the amplitude parameter $\sin^2 2\theta_{13}$ that
are one to two orders of magnitude lower than at a superbeam.

\end{abstract}

\newpage

\section{Introduction}

Measurements~\cite{superk,kam} of the neutrino flux
produced by cosmic ray interactions in the
atmosphere~\cite{atmosflux}
have led to a major breakthrough in our understanding of the
fundamental properties of neutrinos. The early observations
of the atmospheric neutrino interaction rates found a
muon-to-electron event ratio of about
0.6 times the expected ratio. This $\mu/e$ anomaly was
interpreted\cite{old} as evidence for neutrino oscillations with large
amplitude and neutrino mass-squared difference
$\delta m^2_{atm} \sim 10^{-2}$~eV$^2$.
Continued  experimental studies\cite{superk,kam}, especially by the
SuperKamiokande (SuperK)
collaboration, have firmly established that the deviation of the $\mu/e$
ratio from expectation is due to a deficit of muon events. This muon
deficit increases with zenith angle, and hence with path length, and is
consistent with expectations for muon-neutrino oscillations
to some other
neutrino flavor or flavors $(\nu_\mu \rightarrow \nu_x)$
with maximal or near-maximal amplitude and
$\delta m_{\rm atm}^2 \simeq 3.5\times 10^{-3}$~eV$^2$.
In principle $\nu_x$ could be $\nu_e$ (electron-neutrino),
$\nu_\tau$ (tau-neutrino), or $\nu_s$ (sterile neutrino)~\cite{4nu}.
However, the observed
$\nu_e$ flux is in approximate agreement with the
predicted $\nu_e$ flux for all
zenith angles\cite{superk}, which rules out $\nu_\mu \rightarrow \nu_e$
oscillations with large amplitude.
The null results from the CHOOZ and Palo Verde reactor
$\bar{\nu}_e$ disappearance experiments\cite{chooz}
also exclude $\bar{\nu}_e \rightarrow \bar{\nu}_\mu$
oscillations
%(and hence $\bar\nu_\mu \to \bar\nu_e$ if $CPT$ is conserved)
at a mass-squared-difference scale $> 10^{-3}$~eV$^2$ with amplitude $>0.1$.
Furthermore, large amplitude $\nu_\mu \rightarrow \nu_s$ oscillations
at the $\delta m_{\rm atm}^2$ scale are also excluded by SuperK. This is
because $\nu_\mu \rightarrow \nu_s$ oscillations
are expected to be significantly affected by propagation through
matter~\cite{matter,smatter}, causing a distortion in the zenith-angle
distribution at large angles (corresponding to long
path lengths) that is not present in the data~\cite{superk-atmos}.
The zenith-angle distribution observed by SuperK
excludes $\nu_\mu \rightarrow \nu_s$ oscillations of maximal
amplitude at 99\% confidence level\cite{superk-atmos}.
We conclude that, if the oscillation interpretation of the
atmospheric neutrino deficit is correct, the dominant mode must be
$\nu_\mu \rightarrow \nu_\tau$ oscillation,
with the possibility of some smaller amplitude
muon-neutrino oscillations to sterile and/or
electron-neutrinos\cite{fogli-4nu}.

An exotic alternative interpretation\cite{our-nu-decay} of the atmospheric
neutrino
disappearance results is that a neutrino mass-eigenstate,
which is a dominant
component of the $\nu_\mu$ state, decays to a lighter mass-eigenstate and
a Majoron\cite{MajModel}. The first oscillation minimum in $\nu_\mu \to
\nu_\mu$ must be observed or excluded to
differentiate neutrino oscillations from neutrino decays.
Unfortunately, the SuperK neutrino-energy and angular-resolution functions
smear out the characteristic $\nu_\mu$ event rate dip that would
correspond to
the first oscillation minimum, which cannot therefore be resolved.

Progress in establishing neutrino oscillations at the atmospheric
scale is expected in the near future at
accelerator neutrino sources with detectors at medium to long
baselines. The K2K experiment, from KEK to SuperK\cite{k2k}, with a baseline
$L = 250$~km and average neutrino energy $\left< E_\nu \right> =
1.4$~GeV, is
underway. Their preliminary results are in excellent agreement with the
oscillation expectations (27 events are observed, whereas 27 events
would be
expected with oscillations and 40 events for no oscillations\cite{k2k}). The
MINOS experiment from Fermilab to Soudan~\cite{minos} with $L =
730$~km  and $\left< E_\nu\right> = 3.5$~GeV, which
begins  operation in 2003, is expected to resolve the first oscillation
minimum in $\nu_\mu\leftrightarrow\nu_\mu$ and to search for
$\nu_\mu\leftrightarrow\nu_e$ oscillations at the $\delta m_{\rm atm}^2$
scale
with an amplitude sensitivity of $10^{-2}$.
Beginning in 2005, similar physics measurements will be made
by the ICARUS\cite{icarus} and OPERA\cite{opera} experiments
with neutrinos of average energy $\left< E_\nu \right> \simeq 20$~GeV
from CERN detected at $L\simeq 730$~km in the Gran Sasso laboratory.

In addition to the atmospheric neutrino deficit, there are other
possible indications of neutrino oscillations. In particular,
the long-standing deficit of solar
neutrinos~\cite{cleveland,fukuda,solar,GSG}
compared to the Standard
Solar Model (SSM) predictions~\cite{bahcall} is widely interpreted as an
oscillation depletion of the $\nu_e$ flux.
Note that helioseismology and other solar
observations stringently limit uncertainties in the central temperature
of the
sun and other solar model parameters\cite{solartemp}. The $\nu_e$
deficit relative to
prediction is about one-half for the water Cherenkov~\cite{fukuda,solar}
and Gallium experiments~\cite{GSG}, with the Chlorine
experiment~\cite{cleveland} finding a suppression of about one-third.
The latest solar neutrino results from SuperK show an electron recoil
spectrum that is flat in energy, and exhibits no significant day-night or
seasonal variation~\cite{takeuchi}.

An industry has developed to extract the allowed ranges of $\delta
m_{\rm solar}^2$ and $\nu_e$ mixing angles that can account for the
solar neutrino data. The analyses take account of the coherent  scattering
of $\nu_e$ on matter~\cite{matter}, both in the Sun (the MSW
effect~\cite{msw}) and in the Earth~\cite{earth}. These matter effects
can make significant
modifications to vacuum oscillation amplitudes. Until recently, four viable
regions of the parameter space were found in global fits to the data:
\begin{enumerate}
\item[(i)] LMA --- large mixing angle with small matter effects
($\delta m_{\rm solar}^2\approx 10^{-5}$ to $10^{-4}\rm\,eV^2$);
\item[(ii)] SMA --- small mixing angle with large matter effects
($\delta m_{\rm solar}^2\approx 10^{-5}\rm\,eV^2$);
\item[(iii)] LOW --- large-angle mixing with quasi-vacuum amplitude
($\delta m_{\rm solar}^2\approx 10^{-7}\rm\,eV^2$);
\item[(iv)] VO --- large-angle vacuum mixing with small matter effects
($\delta m_{\rm solar}^2\approx 10^{-10}\rm\,eV^2$).
\end{enumerate}

The latest global solar neutrino analysis by the SuperK
collaboration\cite{takeuchi} strongly favors solar $\nu_e$
oscillations to active neutrinos ($\nu_\mu$ and/or $\nu_\tau$)
in the LMA region. A very small area in the LOW region is also allowed at
99\%~C.L., while the SMA and VO regions are rejected at 95\%~C.L. However,
other global analyses disagree that the latter regions are
excluded~\cite{global-3nu}.
Moreover,
%large or small
$\nu_e \rightarrow \nu_s$
solar oscillations may
also still be viable\cite{our-fate,newsmirnov}.
The relative weighting of different experiments (e.g.\ inclusion or
exclusion
of the Cl data) and whether the $^8$B flux is held fixed at its SSM
value or allowed to float in the global fits presumably account in part
for the differences in conclusions. It is expected that the KamLAND reactor
experiment~\cite{kamland} will be able to measure $|\delta m^2_{21}|$
to $\pm10$\% accuracy and $\sin^22\theta_{12}$ to $\pm0.1$
accuracy~\cite{vbkam} if the LMA solar solution is correct.

Finally, the LSND accelerator experiment\cite{LSND} reports
evidence for possible
$\bar\nu_\mu\to\bar\nu_e$ and $\nu_\mu\to\nu_e$ oscillations with very small
amplitude and $\delta m^2_{\rm LSND} \approx 1\rm~eV^2$. If, in
addition to
the LSND observations, the atmospheric and solar effects
are also to be explained by oscillations, then three distinct $\delta m^2$
scales are needed, requiring a sterile neutrino in addition to the three
active neutrino flavors\cite{4nu,fogli-4nu,our-fate,newsmirnov}. The
MiniBooNE experiment at Fermilab\cite{miniboone} is
designed to cover the full region of oscillation parameters indicated by LSND.
%
%We base our present study on three active neutrinos with oscillation
%parameters chosen to describe the atmospheric and solar anomalies for
%the LMA
%solar solution. In this framework we will examine optimal intensity, energy,
%and detector configurations to explore the 3-neutrino oscillation parameter
%space. These considerations for long-baseline experiments are relevant
%even should
%there be additional short-baseline oscillation effects associated with a
%fourth neutrino\cite{our-sbl}.

The principal goals of our analyses are to examine the relative merits of
different neutrino ``superbeam'' scenarios, where we define a neutrino
superbeam
as a conventional neutrino beam generated by $\pi^\pm$ decays, but using a
very intense megawatt(MW)-scale proton source. In particular, we are
interested in the physics reach in medium- and long-baseline experiments
with a neutrino superbeam, and how this reach depends
upon the beam energy, baseline, and the parameters of the neutrino detector.
As representative examples we explicitly consider water Cherenkov,
liquid argon and iron scintillator detectors.
However, we note that there is room for new detector ideas, detector
optimization, and possibly an associated detector R\&D program. Therefore,
results are presented that apply to any detector for
which the effective fiducial mass and background rates can be specified.
Finally, we will discuss the role that neutrino superbeams might
play~\cite{burt,newdick} en route to a neutrino factory~\cite{nuf,our-nuf}.
Our calculations are performed within a  three-neutrino oscillation
framework with the parameters chosen to describe the atmospheric
neutrino deficit and the solar neutrino deficit assuming the LMA
solution. However, our considerations for long-baseline experiments are
relevant even if there are additional short-baseline oscillation effects
associated with a fourth neutrino~\cite{our-sbl}.

The central objective of long-baseline neutrino oscillation experiments
is to determine the
parameters of the neutrino mixing matrix and the magnitudes and signs of
the neutrino mass-squared differences; the signs fix the hierarchy of
the neutrino mass eigenstates~\cite{BGRW}. For three neutrinos $(\nu_e,
\nu_\mu, \nu_\tau)$ the mixing matrix relevant to oscillation phenomena
can be specified by three angles $(\theta_{23}, \theta_{12},
\theta_{13})$ and a phase $\delta$ associated with $CP$-violation [see
Eq.~(\ref{eq:MNS}) below]. There are only two independent mass-squared
differences (e.g.\ $\delta m_{32}^2$ and $\delta m_{21}^2$) for three
neutrinos. The muon-disappearance measurements at SuperK
constrain $\theta_{23}\sim\pi/2$ and $\delta m_{32}^2 \sim 3\times
10^{-3}\rm\,eV^2$. The other parameter that enters at the leading
$\delta m_{32}^2$ oscillation scale is $\theta_{13}$, and its
measurement requires the observation of neutrino appearance in
$\nu_e\to\nu_\mu$, $\nu_\mu\to\nu_e$, or $\nu_e\to\nu_\tau$
oscillations.

In long-baseline experiments, if $\theta_{13}$ is nonzero,
matter effects~\cite{matter,zag,parke} modify the probability for
oscillations involving a $\nu_e$ or $\bar{\nu}_e$ in a way that
can be used to determine the sign of $\delta
m_{32}^2$~\cite{BGRW,our-leading,our-entry,lipari}.
Matter effects give apparent $CP$-violation, but this may be disentangled
from intrinsic $CP$-violation
effects~\cite{BGRW,derujula,donini,dick,freund,ayres,albright,campanelli,%
koike,yasuda}
at optimally chosen baselines. Matter can also modify the effects of
intrinsic $CP$ or $T$ violation~\cite{alter}. Intrinsic $CP$ violation
may also be studied at short baselines where matter effects are relatively
small~\cite{arafune,wakaizumi}. $CP$-violating effects enter only for
values of  $L/E_\nu$
where oscillations associated with the subleading $\delta m_{21}^2$
become significant~\cite{BPW,P}. The most challenging goal of
accelerator-based neutrino oscillation experiments is to detect,
or place stringent limits on, $CP$ violation in the lepton sector.
We will address the extent to which this may
be possible with conventional superbeams.

\section{Three-Neutrino Formalism}
\label{sec:theory}

The flavor eigenstates $\nu_\alpha\ (\alpha = e, \mu, \tau)$ are related to
the mass eigenstates $\nu_j\ (j = 1, 2, 3)$ in vacuum by
\begin{equation}
\nu_\alpha = \sum_j U_{\alpha j} \nu_j \;,
\end{equation}
where $U$ is a unitary $3\times3$ mixing matrix. The propagation of
neutrinos through matter is described by the evolution
equation~\cite{matter,kuo}
\begin{equation}
i{d\nu_\alpha\over dx} = \sum_\beta \left( \sum_j U_{\alpha j}
U^*_{\beta j}
{m^2_j\over 2E_\nu} + {A\over 2E_\nu} \delta_{\alpha e} \delta_{\beta e}
\right) \nu_\beta \;,
\label{eq:evolution}
\end{equation}
where $x = ct$ and $A/2E_\nu$ is the amplitude for coherent forward
charged-current $\nu_e$ scattering on electrons,
\begin{equation}
A = 2\sqrt 2\, G_F \, Y_e \, \rho\, E_\nu = 1.52\times10^{-4}\,{\rm eV^2}
\times Y_e \, \rho\,({\rm g/cm^3}) \times E_\nu\,(\rm GeV) \;,
\label{eq:A}
\end{equation}
where $Y_e(x)$ is the electron fraction and $\rho(x)$ is the matter density.
In the Earth's crust the average density is typically 3--4~gm/cm$^3$
and $Y_e\simeq 0.5$. The propagation equations can be
re-expressed in terms of mass-squared differences:
\begin{equation}
i{d\nu_\alpha\over dx} = \sum_\beta {1\over 2E_\nu} \left( \delta m_{31}^2
U_{\alpha3} U_{\beta3}^* + \delta m_{21}^2 U_{\alpha2} U_{\beta2}^* + A
\delta_{\alpha e} \delta_{\beta e}\right) \;,
\end{equation}
where $\delta m_{jk}^2 = m_j^2 - m_k^2$. We assume $|\delta m^2_{21}|
\ll |\delta m^2_{32}|$, and that the sign of $\delta m^2_{32}$ can be
either positive or negative, corresponding to the case where the most
widely separated mass eignstate is either above or below, respectively,
the other two mass eigenstates. Thus the sign of $\delta m^2_{32}$
determines the ordering of the neutrino masses.
The evolution equations can be solved numerically taking into account the
dependence of the density on depth using the density profile from the
preliminary reference earth model\cite{PREM}. We integrate the equations
numerically along the neutrino path using a Runge-Kutta
method. The step size at each point along the path is taken to be 1\% of
the shortest oscillation wavelength given by the two scales
$\delta m^2_{32}$ and $A$.

It is instructive to examine analytic expressions for the vacuum
probabilities. We introduce the notation
\begin{equation}
\Delta _{jk} \equiv \delta m_{jk}^2 L/4E_\nu = 1.27 (\delta m_{jk}^2/{\rm eV^2}
(L/{\rm km}) ({\rm GeV}/E_\nu) \;.
\end{equation}
The vacuum probabilities are then given by
\begin{eqnarray}
P(\nu_\alpha\to\nu_\beta) &=& -4\Re (U_{\alpha 2} U_{\alpha3}^* U_{\beta2}^*
U_{\beta3}) \sin^2\Delta_{32} - 4\Re (U_{\alpha 1} U_{\alpha3}^*
U_{\beta1}^*
U_{\beta3}) \sin^2\Delta_{31}\nonumber\\
&& -4\Re (U_{\alpha 1} U_{\alpha2}^* U_{\beta1}^* U_{\beta2})
\sin^2\Delta_{21} \pm 2JS \;,
\end{eqnarray}
where $J$ is the $CP$-violating invariant\cite{keung,jarlskog},
\begin{equation}
J = \Im(U_{e2} U_{e3}^* U_{\mu2}^* U_{\mu3}) \;,
\end{equation}
and $S$ is the associated dependence on $L$ and $E_\nu$,
\begin{equation}
S = \sin2\Delta_{21}  + \sin2\Delta_{32} - \sin2\Delta_{31} \; .
\label{eq:s}
\end{equation}
The mixing matrix can be specified by 3 mixing angles ($\theta_{32},
\theta_{12}, \theta_{13}$) and a $CP$-violating phase ($\delta$). We adopt the
parameterization
\begin{equation}
U
= \left( \begin{array}{ccc}
  c_{13} c_{12}       & c_{13} s_{12}  & s_{13} e^{-i\delta} \\
- c_{23} s_{12} - s_{13} s_{23} c_{12} e^{i\delta}
& c_{23} c_{12} - s_{13} s_{23} s_{12} e^{i\delta}
& c_{13} s_{23} \\
    s_{23} s_{12} - s_{13} c_{23} c_{12} e^{i\delta}
& - s_{23} c_{12} - s_{13} c_{23} s_{12} e^{i\delta}
& c_{13} c_{23} \\
\end{array} \right) \,,
\label{eq:MNS}
\end{equation}
where $c_{jk} \equiv \cos\theta_{jk}$ and $s_{jk} \equiv
\sin\theta_{jk}$. We
can restrict the angles to the first quadrant, $0\le \theta_{ij} \le \pi/4$,
with $\delta$ in the range $-\pi \le \delta \le \pi$.
In this parameterization $J$ is given by
\begin{equation}
J = s_{13} c_{13}^2 s_{12} c_{12} s_{23} c_{23} s_\delta =
{1\over8} \sin2\theta_{23} \sin2\theta_{12} \sin2\theta_{13} c_{13} s_\delta
\; .
\end{equation}
For convenience we also define
\begin{equation}
K =  s_{13} c_{13}^2 s_{12} c_{12} s_{23} c_{23} c_\delta =
{1\over8} \sin2\theta_{23} \sin2\theta_{12} \sin2\theta_{13} c_{13} c_\delta
\; .
\end{equation}
Then the vacuum appearance probabilities are given by
\begin{eqnarray}
P(\nu_\mu\to\nu_e) &=& \left[ s_{23}^2 s_{12}^2 \sin^22\theta_{13} -
4K \right] \sin^2\Delta_{32}
+ \left[ s_{23}^2 c_{12}^2 \sin^22\theta_{13} + 4K \right]
\sin^2\Delta_{31} \nonumber\\
&&{}+ \left[ c_{13}^2 (c_{23}^2 - s_{13}^2 s_{23}^2) \sin^22\theta_{12}
+ 4K\cos2\theta_{12}  \right] \sin^2\Delta_{21} + 2JS
\label{eq:Pme}\\
P(\nu_\mu\to\nu_\tau) &=&
\left[ c_{13}^2(c_{12}^2-s_{12}^2s_{13}^2)\sin^22\theta_{23}
+ 4K\cos\theta_{23} \right] \sin^2\Delta_{32}
\nonumber\\
&&{}+ \left[ c_{13}^2(s_{12}^2-c_{12}^2s_{13}^2)\sin^22\theta_{23}
- 4K\cos\theta_{23} \right] \sin^2\Delta_{31}
\nonumber\\
&&{}+2\left[ \sin^22\theta_{23} \left( s_{13}^2-s_{12}^2c_{12}^2
(1+s_{13}^2)^2 \right)
+s_{13}^2\sin^22\theta_{12}(1+\sin^22\theta_{23}s_\delta^2) \right.
\nonumber\\
&&{}+\left.
\sin2\theta_{12}\cos\theta_{12}\sin\theta_{23}\cos\theta_{23}s_{13}
c_\delta (1+s_{13}^2) \right] \sin^2\Delta_{21} -2JS
\label{eq:Pmt}\\
P(\nu_e\to\nu_\tau) &=& \left[ c_{23}^2 s_{12}^2 \sin^22\theta_{13} +
4K \right] \sin^2\Delta_{32}
+ \left[ c_{23}^2 c_{12}^2 \sin^22\theta_{13} - 4K  \right]
\sin^2\Delta_{31} \nonumber\\
&&{}+ \left[ c_{13}^2 (s_{23}^2 - s_{13}^2 c_{23}^2) \sin^22\theta_{12}
- 4K\cos2\theta_{12}  \right] \sin^2\Delta_{21} +2JS
\label{eq:Pet}
\end{eqnarray}
The corresponding probabilities $P(\bar\nu_\alpha\to\bar\nu_\beta)$
can be obtained by reversing the sign of $\delta$ in the above
formulas (only the $JS$ term changes sign in each case).
The probabilities for $\bar\nu_\beta\to\bar\nu_\alpha$ are the same as
those for $\nu_\alpha\to\nu_\beta$, assuming $CPT$ invariance.
Tests of $CPT$ non-invariance are
important~\cite{coleman,our-cpt,murayama}
but beyond the scope of the present analysis.
The $\Delta_{ij}$ are not independent, and can be expressed in terms of
$\Delta_{atm} \equiv \Delta_{32}$ and $\Delta_{sun} \equiv \Delta_{21}$.
Then $\Delta_{31} = \Delta_{atm} + \Delta_{sun}$ and
\begin{eqnarray}
\sin^2\Delta_{31} &=&
\sin^2\Delta_{atm} + \sin^2\Delta_{sun}\cos2\Delta_{atm}
+{1\over2}\sin2\Delta_{sun}\sin2\Delta_{atm} \; ,
\label{eq:delta31}\\
S &=& 2\left( \sin2\Delta_{sun} \sin^2\Delta_{atm}
+ \sin2\Delta_{atm} \sin^2\Delta_{sun} \right) \; .
\label{eq:s2}
\end{eqnarray}

Since Eqs.~(\ref{eq:Pme})--(\ref{eq:s2}) in their exact form are
somewhat impenetrable, we make a few simplifying assumptions to
illustrate their typical consequences. First, it is advantageous in
long-baseline experiments to operate at an $L/E_\nu$ value such that the
leading oscillation is nearly maximal, i.e. $\Delta_{atm} \simeq
\pi/2$. Since $\delta m^2_{sun} \ll \delta m^2_{atm}$,
$\Delta_{sun} \ll 1$ and to a good approximation we can ignore terms
involving $\sin^2\Delta_{sun}$. Also, since $\theta_{13}$ is
already constrained by experiment to be small,
for the terms involving $\Delta_{sun}$ we retain only the leading
terms in $\theta_{13}$. Second,
at the value $\Delta_{atm} \simeq \pi/2$ for which the leading
oscillation is best measured, $\sin2\Delta_{atm} \simeq 0$. Even if
$\Delta_{atm}$ is not close to $\pi/2$ for all neutrino energies in the
beam, an averaging over the energy spectrum will suppress $\sin
2\Delta_{atm}$ if the neutrinos at the middle of the spectrum have
$\Delta_{atm} \simeq \pi/2$. With the above approximations the
vacuum oscillation probabilities simplify to
\begin{eqnarray}
P(\nu_\mu\to\nu_e) &\simeq&
\sin^2\Delta_{atm}
\left( s_{23}^2 \sin^22\theta_{13} + 4J\sin2\Delta_{sun} \right)
\label{eq:Pme2}\\
P(\nu_\mu\to\nu_\tau) &\simeq&
\sin^2\Delta_{atm}
\left( \sin^22\theta_{23} - 4J\sin2\Delta_{sun} \right) \;
\label{eq:Pmt2}\\
P(\nu_e\to\nu_\tau) &\simeq&
\sin^2\Delta_{atm}
\left( c_{23}^2 \sin^22\theta_{13} + 4J\sin2\Delta_{sun} \right)
\label{eq:Pet2}
\end{eqnarray}
It is interesting to compare the relative sizes of the leading
$CP$-violating ($CPV$)
and $CP$-conserving ($CPC$) terms in the $\nu_\mu \to
\nu_e$ oscillation probability:
\begin{equation}
{CPV\over CPC} \simeq
{4J\sin2\Delta_{sun}\over s_{23}^2\sin^22\theta_{13}}
\simeq \left( {\sin2\theta_{12}\sin2\theta_{23}\over2s_{23}^2} \right)
\left( {\Delta_{sun} \sin\delta \over \theta_{13}} \right) \; .
\label{eq:cpv}
\end{equation}
For the standard three-neutrino solution to the solar and atmospheric
data with large-angle mixing in the solar sector, the first fraction on
the right-hand side of Eq.~(\ref{eq:cpv}) is of order unity, and the
relative size of the $CPV$ term is
\begin{equation}
{CPV\over CPC} \sim {\Delta_{sun}\sin\delta \over \theta_{13}} \; .
\label{eq:cpv2}
\end{equation}

As an example, with $\delta m^2_{sun} \simeq 1\times10^{-4}$~eV$^2$ and
$L/E_\nu \simeq 300$~km/GeV, $\Delta_{sun} \simeq 0.04$; then with
$\delta = \pi/2$ and $\sin^22\theta_{13} = 0.1$ (its maximum allowed
value), the $CPV$ term is about 25\% of the $CPC$ term. Smaller values
of $\delta m^2_{21}$ or $\sin\delta$ decrease the ratio in
Eq.~(\ref{eq:cpv2}); smaller values of $\theta_{13}$ increase it.

While smaller values of $\theta_{13}$ give a larger relative $CPV$ term,
they will also reduce the overall $\nu_\mu \to \nu_e$ event rate since
the $CPC$ term is
proportional to $\sin^22\theta_{13}$. The $CPV$ effect may be hard
to measure if the event rate is low (due to insufficient flux or a small
detector). Because the number of $CPC$ events is proportional to
$\sin^22\theta_{13}$, the statistical uncertainty on the $CPC$ event rate
is proportional to $\theta_{13}$ for small $\theta_{13}$ and Gaussian
statistics. Since the number
of $CPV$ events is also proportional to $\theta_{13}$, the size of the
$CPV$ signal relative to the statistical uncertainties does not decrease
as $\theta_{13}$ becomes smaller.
%Roughly speaking, as
%long as the event rate is high enough for there to be a statistically
%relevant number of $CPV$ events (e.g., ten or more), then $CPV$ may be
%detectable, regardless of the size of $\theta_{13}$
Therefore, a priori it does not follow that small $\theta_{13}$
automatically makes $CPV$ undetectable~\cite{marciano-conf}.

On the other hand, even if the event rate is high enough to overcome the
statistical uncertainties on the signal, backgrounds will limit the
ability to measure $CPV$. Background considerations place an effective
lower bound on the
values of $\sin^22\theta_{13}$ for which a $CPV$ search can be made.
This can be quantified by noting
that the ratio of the number of $CPV$ events $N_{CPV}$ to the
uncertainty due to the background is $N_{CPV}/\sqrt{f_B N_0}$ (assuming
Gaussian statistics), where $N_0$ is the number of events without
oscillations and $f_B$ is the background fraction. $N_{CPV}$ can be
expressed as the $CPV$ part of the oscillation probability times $N_0$.
Using the expression for the probability in Eq.~(\ref{eq:Pme2}), it
follows that a $3\sigma$ $CPV$ effect in the $\nu_\mu \to \nu_e$
channel requires
\begin{equation}
\sin^22\theta_{13} \ge
{9\over\Delta^2_{sun} \; \sin^2\delta} \; {f_B\over N_0} \; .
\label{eq:th13min}
\end{equation}
For $\delta =\pi/2$ and $\Delta_{sun} = 0.03$, a typical experiment with
$f_B = 0.01$ and $N_0 = 10^4$ can detect $CPV$ for $\sin^22\theta_{13}
\ge 0.01$. The detailed calculations in Sec.~\ref{sec:cpv} confirm this
approximate result.

The preceding discussion applies only when the corrections due to matter
are not large, generally when $L$ is small compared to the Earth's
radius. Reference~\cite{arafune} gives approximate expressions for the
probabilities when the matter corrections are small but not negligible.
However, the most striking matter effects occur when the matter
corrections are large and the expansions of Ref.~\cite{arafune}
are no longer valid (see, e.g., the plots of oscillation probabilities
in matter given in Ref.~\cite{shrock}).

Some of the qualitative properties of
neutrino oscillations in matter can be determined by
considering only the leading oscillation and assuming a constant
density. There is an effective mixing angle in matter defined by
\begin{equation}
\sin^2 2\theta_{13}^m = {\sin^2 2\theta_{13}\over
\left({A\over\delta m^2} - \cos 2\theta_{13} \right)^2
+ \sin^2 2\theta_{13}} \,. \label{eq:sin}
\end{equation}
where $A$ is given in Eq.~(\ref{eq:A}).
The oscillation probabilities in the leading oscillation
approximation for constant density are~\cite{our-leading,pantaleone}
\begin{eqnarray}
P(\nu_\mu \to \nu_e) &=&
s_{23}^2 \sin^2 2\theta_{13}^m \sin^2\Delta_{32}^m \,,
\nonumber\\
P(\nu_\mu \to \nu_\tau) &=& \sin^2 2\theta_{23} \left[
 (\sin\theta_{13}^m)^2 \sin^2\Delta_{21}^m
+(\cos\theta_{13}^m)^2 \sin^2\Delta_{31}^m
-(\sin\theta_{13}^m\cos\theta_{13}^m)^2 \sin^2\Delta_{32}^m \right]
\,, \nonumber\\
P(\nu_e \to \nu_\tau) &=&
c_{23}^2 \sin^2 2\theta_{13}^m \sin^2\Delta_{32}^m \,,
\label{eq:probs}
\end{eqnarray}
where the oscillation arguments are
\begin{equation}
\Delta_{32}^m = \Delta_{atm} S \,,\qquad
\Delta_{31}^m =
\Delta_{atm} {1\over2} \left[ 1+{A\over\delta m^2_{atm}}+S \right]
\,, \qquad
\Delta_{21}^m =
\Delta_{atm} {1\over2} \left[ 1+{A\over\delta m^2_{atm}}-S \right]
\,, \label{eq:arg}
\end{equation}
and
\begin{equation}
S \equiv
\sqrt{ \left( {A\over\delta m^2_{atm}}-\cos2\theta_{13} \right)^2
+ \sin^2 2\theta_{13}} \,.
\label{eq:S}
\end{equation}
The $\Delta_{21}^m$ term in $P(\nu_\mu\to\nu_\tau)$ must be
retained here because it is not necessarily negligible compared to
$\Delta_{31}^m$, due to matter effects. The expressions for
antineutrinos may be generated by changing the sign of $A$.

In Eq.~(\ref{eq:sin}) there is a resonant
enhancement of $\nu_\mu \to \nu_e$ oscillations when $A \simeq
\delta m^2_{atm} \cos2\theta_{13}$ ($A  \simeq - \delta m^2_{atm}
\cos2\theta_{13}$ for antineutrinos). This occurs for
neutrinos when $\delta m^2_{atm} > 0$ and for antineutrinos when
$\delta m^2_{atm} <0$. On resonance, there is a suppression for
antineutrinos (neutrinos) when $\delta m^2_{atm} > 0$ ($\delta m^2_{atm}
< 0$). This enhancement of one channel and suppression of the other
then gives a fake $CP$ violation due to matter effects.

In the event that the contribution of the sub-leading oscillation is not
negligible, the true
$CPV$ effects due to $\delta$ also enter, but they may be masked by
matter effects.  Numerical calculations~\cite{our-entry} show
that for distances larger than 2000~km matter effects dominate the
true $CPV$ for $\sin^22\theta_{13} > 0.001$ and vice versa for
$\sin^22\theta_{13} < 0.001$.

As long as $\sin^22\theta_{13}$ is not too small, one approach is to have
$L$ large enough so that the dominant $CPV$ effect is from matter and
the sign of $\delta m^2_{atm}$ is clearly determinable; then the true
$CPV$ effect can be extracted by considering deviations from the
$CP$-conserving predictions~\cite{our-entry}. With large  $L$,
the neutrino energies must be high enough that
$\Delta_{atm} \simeq \pi/2$ (e.g., $L \sim 3000$~km requires $E_\nu \sim
10$~GeV). An alternative approach is to have short $L$ where the
matter effects are relatively small~\cite{arafune}; this usually
requires a smaller $E_\nu$ to have $\Delta_{atm}$  of order
$\pi/2$ (e.g., $L \sim 300$~km and $E_\nu \sim 1$~GeV). We will study both
of these possibilities in this paper.

For $\sin^22\theta_{13} \le 0.001$, the matter effect is similar in size
or smaller than the true $CPV$ effect, and it may not be possible to
distinguish between large
intrinsic $CPV$ with very small $\theta_{13}$ from no intrinsic
$CPV$ with a moderate-sized $\theta_{13}$. Even in experiments at short
distances (where the matter effect is small) the number of appearance
events may be too small relative to the background to have a
statistically significant
difference between the neutrino and antineutrino oscillation probabilities.
The
existence of intrinsic $CP$ violation may be very difficult to determine
in this case.

\section{Conventional neutrino beams}
\label{sec:nu-beams}

Conventional neutrino beams are produced using a pion decay channel.
If the pions are charge-sign selected so that only positive (negative)
particles are within the channel, the pion decays
$\pi^+ \rightarrow \mu^+ \nu_\mu$
($\pi^- \rightarrow \mu^- \bar{\nu}_\mu$)
will produce a beam of muon neutrinos (antineutrinos).
The beams will also contain small components of $\nu_e$ and
$\bar{\nu}_e$ from kaon and muon decays.
For a positive beam, the dominant decays that contribute to the
$\nu_e$ component are
$K^+ \rightarrow \pi^0 e^+ \nu_e$ and
$\mu^+ \rightarrow e^+ \nu_e \bar{\nu}_\mu$.
If the pion beam has not been charge-sign selected there will also be
a contribution from $K^0_L \rightarrow \pi^\pm e^\mp \nu_e$ decays.
The $\nu_e + \bar{\nu}_e$ ``contamination" can be
minimized using beam optics that disfavor decays occurring close
to the target [note: $\tau(K^\pm) \sim 0.5 \tau(\pi^\pm)$], and choosing
a short decay channel to reduce the contribution from muon decays.
These strategies enhance the flavor purity of the beam, but reduce the
beam flux.
Depending on the beamline design, the resulting $\nu_e + \bar{\nu}_e$
contamination is typically at the few parts in 100 to a few parts in
1000 level.
The intrinsic $\nu_e$ component in the beam produces a background that
must be subtracted in a $\nu_\mu \rightarrow \nu_e$ oscillation search.
Ultimately, the systematic uncertainty associated
with the background subtraction will degrade the sensitivity of the
oscillation measurement.

To maximize the neutrino flux in the forward direction it is desirable
that the pion beam divergence is small within the decay channel. The
required radial focusing can be provided by a quadrupole
channel and/or magnetic horns. The beamline optics (dipoles, horns,
and quadrupoles) determine the peak pion energy and energy spread
within the decay channel, and hence determine the neutrino spectrum.
If the optics are designed to accept a large pion momentum spread
the resulting wide band beam (WBB) will contain a large neutrino flux
with a broad energy spectrum. If the optics are designed to accept a
smaller pion momentum spread, the resulting narrow band beam (NBB)
will have a narrower energy spread, but a smaller flux.

\subsection{Detectors and backgrounds}

We are primarily interested in searching for, and measuring,
$\nu_\mu \rightarrow \nu_e$
and $\bar{\nu}_\mu \rightarrow \bar{\nu}_e$ oscillations.
The experimental signature for these oscillation modes is the
appearance of an energetic electron or positron in a
charged-current (CC) event.
The electron
must be separated from the hadronic remnants produced by the
fragmenting nucleon.
Backgrounds can arise from (i) energetic neutral pions that are
produced in neutral-current
(NC) interactions, and subsequently fake a
prompt-electron signature, (ii) energetic neutral pions that are
produced in CC interactions in which the muon is undetected, and
the $\pi^0$ fakes an electron, (iii) charm production and
semileptonic decay, (iv) $\nu_\mu \rightarrow \nu_\tau$
oscillations followed by  decay of the tau-lepton to an electron.
Backgrounds (iii) and (iv) can be suppressed using a low-energy
neutrino beam.

Background (i) is potentially the most dangerous
since leading $\pi^0$ production in NC events is
not uncommon. Indeed, in a recent study~\cite{jhf_loi,otherwater}
using a low energy $\nu_\mu$ beam
it has been shown that in a water Cherenkov detector (e.g. SuperK)
it is difficult to
reduce this background to a level below $\cal O$(3\%) of the CC rate.
A liquid argon detector is believed to provide much better
$\pi^0$-electron discrimination, and will perhaps enable the $\pi^0$ background
to be reduced to $\cal O$(0.1\%) of the CC rate~\cite{private_comm}.
Based on these considerations there are two different detector
strategies. We can choose a water Cherenkov detector,
enabling us to maximize the detector mass, and
hence the event statistics, but obliging us to tolerate
a significant background
from $\pi^0$ production in NC events. Alternatively, we can
choose a detector technology that highly suppresses the $\pi^0$ background,
but this will oblige us to use a smaller fiducial mass, and hence lower
event statistics.

For a given choice of beamline design, baseline, and
detector parameters, the experimental $\nu_\mu \rightarrow \nu_e$
and $\bar{\nu}_\mu \rightarrow \bar{\nu}_e$
sensitivities can be calculated.
It is useful to define some representative scenarios,
characterized by
(a) the parameters of the primary proton beam
incident on the pion production target,
(b) the data sample size $D$ (kt-years), defined as the product of the
detector fiducial mass, the efficiency of the signal selection requirements,
and the number of years of data taking,
(c) the background fraction $f_B$ defined as the background rate divided by
the CC rate for events that survive the signal selection requirements,
and (d) the fractional systematic uncertainty $\sigma_{f_B}/f_B$ on the
predicted $f_B$.

We will consider neutrino superbeams that can be produced with
MW-scale proton beams at low energy ($E_\nu \sim 1$~GeV) at the
proposed Japan Hadron Facility (JHF) and at high energy ($E_\nu \ge
3$~GeV) at laboratories with high-energy proton drivers
that might be upgraded to produce these superbeams (BNL, CERN,
DESY, and Fermilab). For the sake of definiteness,
in the following we will restrict our considerations to
two explicit MW-scale primary proton beams.
First, we will consider the 0.77~MW beam at the 50~GeV proton
synchrotron of the proposed JHF, and a 4~MW
upgrade which we refer to as SJHF (Superbeam JHF).
Second, we consider a 1.6~MW proton driver upgrade that is under study
at Fermilab.
With this new proton driver, and modest upgrades to the 120~GeV
Fermilab Main Injector (MI), it is possible to increase the beam current
within the MI by a factor of four,  and hence increase the intensity
of the NuMI beam by a factor of four, which we refer to as SNuMI (Superbeam
NuMI). Higher beam intensities are precluded
by space-charge limitations in the MI~\cite{weiren}.
For both the SJHF and SNuMI cases we will assume a run plan in which
there is 3~years of data taking with a neutrino beam followed by
6~years of data taking with an antineutrino beam.
In principle the high-energy superbeams could be produced at
any of the present laboratories with high-energy proton drivers.

We define three aggressive detector scenarios, which are summarized in
Table~\ref{tab:xx}:
\begin{description}
\item{{\bf Scenario $A$}, which might be realized with a liquid argon detector.}
We choose a 30~kt fiducial mass, which
has been considered previously for a neutrino factory detector.
We assume tight
selection requirements are used to suppress the $\pi^0$ background, and
take the signal efficiency to be 0.5.
This will result in $D = 45$~kt-years for neutrino running and
$90$~kt-years for antineutrino running. We assume that backgrounds from
$\pi^0$ events contribute 0.001 to $f_B$~\cite{mario}, and the
$\nu_e$ contamination
in the beam contributes 0.003 to $f_B$ in neutrino running, and
$0.005$ to $f_B$ in antineutrino running. We neglect all other backgrounds.
Hence, $f_B = 0.004$ (0.006)
for neutrino (antineutrino) running. Finally, we will assume that we know
the background rate with a precision of 10\% ($\sigma_{f_B}/f_B = 0.1$).

\item{{\bf Scenario $F$}, which might be realized with a fine-grain iron
sampling
calorimeter.}
We choose a 10~kt fiducial mass, which is a factor of 10 larger than the
THESEUS detector~\cite{theseus}. Since this fiducial mass is smaller
than for the alternative
scenarios we are considering, we will assume that the selection cuts are
not tight, and that the selection efficiency is 0.9. This will result
in $D = 27$~kt-years for neutrino running and
$54$~kt-years for antineutrino running.
 We assume that backgrounds from
$\pi^0$ events contribute 0.01 to $f_B$, and the $\nu_e$ contamination
in the beam contributes 0.003 to $f_B$ in neutrino running, and
$0.005$ to $f_B$ in antineutrino running. We neglect all other backgrounds.
Hence, $f_B = 0.013$ (0.015)
for neutrino (antineutrino) running. Finally, we will assume that we know
the background rate with a precision of 10\% ($\sigma_{f_B}/f_B = 0.1$).

\item{{\bf Scenario $W$}, which might be realized for a low-energy beam
with a water Cherenkov detector.}
We choose a 220~kt fiducial mass, a factor of 10 larger than
the SuperK detector. Guided by the study described in Ref.~\cite{jhf_loi}
we will assume the
selection requirements used to suppress the $\pi^0$ background result
in a signal efficiency of 0.68.
This will result in $D = 450$~kt-years for neutrino running and
$900$~kt-years for antineutrino running. We assume that backgrounds from
$\pi^0$ events dominate, and set $f_B = 0.02$. Note that a detailed
detector simulation has obtained $f_B = 0.03$ for a water Cherenkov
detector at a low energy NBB at the JHF~\cite{jhf_loi}. With further
optimization the choice $f_B = 0.02$ might therefore be realizable at
low energy, but for higher energy ($> 1$~GeV) neutrino beams the
rejection against the $\pi^0$ background is expected to be much worse.
Hence, a
new detector technology might be required for this scenario to make
sense at high energies. Finally, we will assume that we know
the background rate with a precision of 10\% ($\sigma_{f_B}/f_B = 0.1$).
\end{description}

Scenarios $A$, $F$, and $W$ are very aggressive, and may or may not be
realizable in practice. In the following we will explore the
oscillation sensitivity as a function of $D$, $f_B$, $\sigma_{f_B}/f_B$,
baseline, and neutrino beam energy. Scenario $F$ is clearly inferior to
scenario $A$, which has larger $D$ and smaller $f_B$. Therefore, in
the following we will
not discuss the scenario $F$ physics potential in detail, but we will
indicate the scenario $F$ physics potential on several relevant figures.
We will use scenarios $A$ and $W$ extensively to
illustrate the physics potential of upgraded conventional
neutrino beams, and facilitate a discussion of the challenges
involved in probing small values of $\sin^2 2\theta_{13}$.

\section{$\sin^2 2\theta_{13}$ reach}
\label{sec:reach}

For a given neutrino beam, baseline, and detector, we wish to
calculate the resulting $\sin^2 2\theta_{13}$ reach, which we define
as the value of $\sin^2 2\theta_{13}$ that would result in a
$\nu_\mu \rightarrow \nu_e$ or
$\bar{\nu}_\mu \rightarrow \bar{\nu}_e$ signal that
is 3 standard deviations above the background. In our analysis
we will take into account the Poisson statistical
uncertainties on the numbers of signal and background events,
and the systematic uncertainty on the background subtraction.
Our prescription for
determining the $\sin^22\theta_{13}$ reach is given in the appendix.
In the following, unless otherwise stated, the $\sin^2 2\theta_{13}$
reaches are calculated setting the sub--leading oscillation
parameters $\sin^2 2\theta_{12} = 0.8$ and
$\delta m^2_{21} = 10^{-5}$~eV$^2$. The small $\delta m^2_{21}$
effectively switches off contributions from the sub--leading scale.
Larger values of $\delta m^2_{21}$ can yield contributions to
the $\nu_\mu \rightarrow \nu_e$ signal, but these contributions are
not important unless $\sin^2 2\theta_{13}$ is significantly less than
$10^{-3}$.

\subsection{Sensitivity at the Japan Hadron Facility}

The Japan Hadron Facility working group has recently
investigated~\cite{jhf_loi} the $\nu_\mu \rightarrow \nu_e$ oscillation
sensitivity attainable at the proposed 0.77~MW 50~GeV proton synchrotron
in Japan, using a 295~km baseline together with the SuperK detector. In
their study they considered a variety of low energy
($\left< E_\nu \right> \sim 1$~GeV)
WBB and NBB beamline designs.  The resulting experimental
scenario is similar to the one later considered in Ref.~\cite{burt}. The
conclusions from the study were: (i)~With a water Cherenkov detector the
sensitivity is limited by the $\pi^0$ background produced in NC
events. To minimize the background a NBB must be used, since in a WBB
the high energy tail will be the dominant source of background events.
(ii)~With an effective $\nu_\mu\to\nu_e$ oscillation amplitude of
$\sin^2 2\theta_{\mu e} = 0.05$ at an oscillation scale of $\delta
m^2_{32} = 0.003$~eV$^2$, the best set of selection requirements
identified in the study yielded $f_B = 0.03$, and a signal:background
ratio of 1:1 with 12.3 signal events per year in the SuperK detector.  A
10\% systematic uncertainty on the background rate was assumed.  If no
signal is observed after 5 years of running the expected limit would be
$\sin^2 2\theta_{\mu e} < 0.01$ at 90\% C.L., which corresponds to a
$\sin^2 2\theta_{13}$ reach of 0.05.  (iii)~With a detector of mass 20
$\times$ SuperK, after 5~years running the resulting limit in the
absence of a signal would be $\sin^2 2\theta_{\mu e} < 0.003$ at 90\%
C.L., which corresponds to a $\sin^2 2\theta_{13}$ reach of 0.01.  (iv)~The  
energy distribution of the background events is similar to the
corresponding distribution for the signal. Hence, the $\nu_\mu
\rightarrow \nu_e$ search is essentially a counting experiment.  Using
these JHF study results, we find that the appropriate values to use in
evaluating the $\sin^2 2\theta_{13}$ reach for the JHF to SuperK
experiment are
$f_B = 0.03$, $\sigma_{f_B}/f_B = 0.1$, and $D = 75$~kt-years (SuperK
with 5~years exposure and a signal efficiency of 68\%).  With these
values, our statistical treatment recovers the 90\%~C.L. results
presented in the JHF report~\cite{jhf_loi}. In our calculations for SJHF
we used the neutrino intraction rates presented in Ref.~\cite{jhf_loi}.

\begin{figure}
\centering\leavevmode
\epsfxsize=6.0in\epsffile{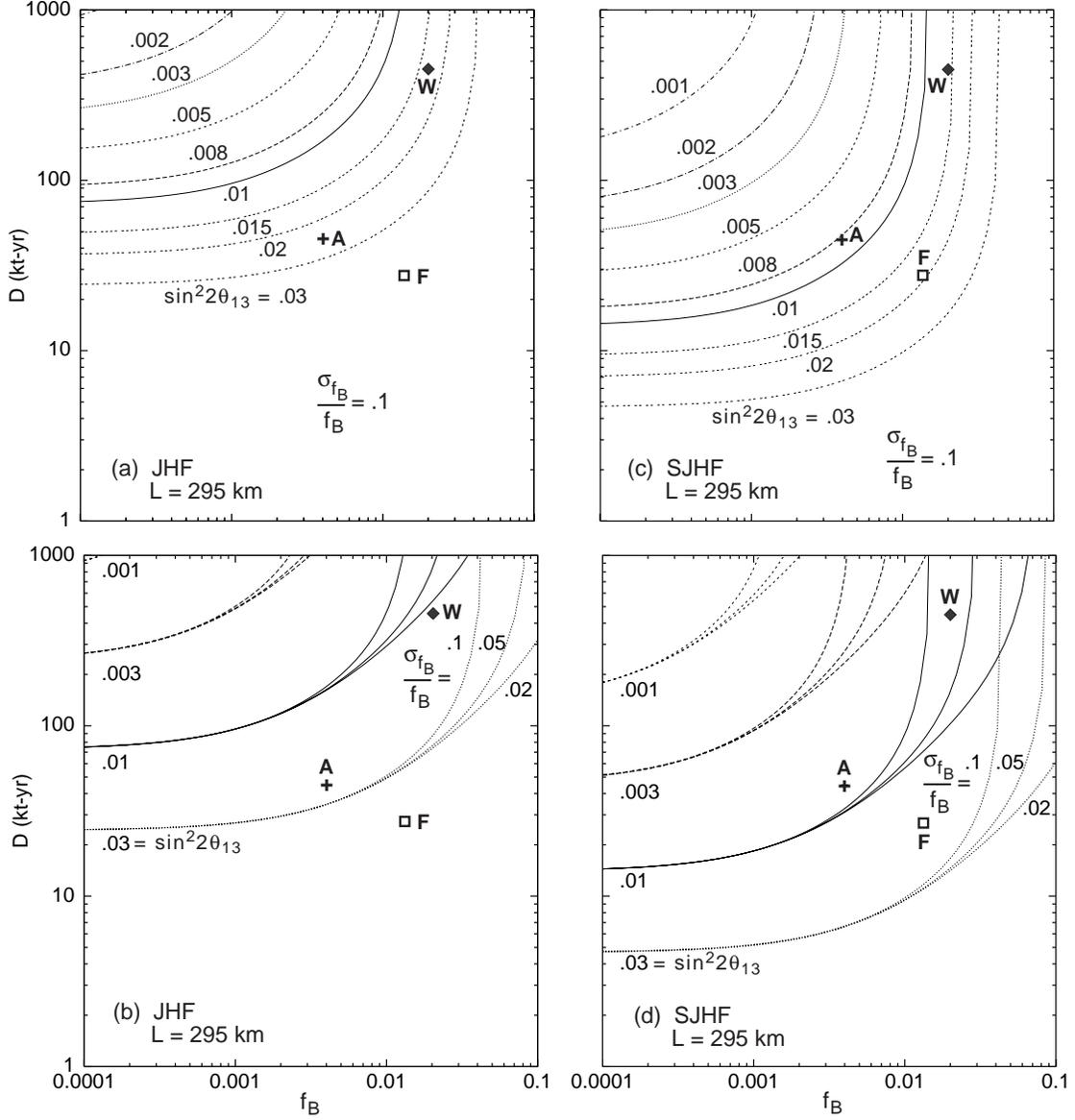}

\medskip
\vspace{0.5cm}
\caption[]{Contours of constant $\sin^22\theta_{13}$ reach that
correspond to a $\nu_e\to \nu_\mu$ signal that is 3 standard deviations
above the background. The contours are shown in the $(D, f_B)$-plane,
where $D$ is the data-sample size and $f_B$ the background rate divided
by the total CC rate.
The contours are shown for
the 0.77~MW (left-hand plots) and 4.0~MW (right-hand plots)
JHF scenarios with
$L = 295$~km. The top panels show curves for $\sigma_{f_B}/f_{B} =
0.1$, while the bottom panels show curves for $\sigma_{f_B}/f_{B} =
0.1$, $0.05$, and $0.02$. The positions corresponding to the
three standard detector scenarios defined in Table~\ref{tab:xx}
are indicated.}
\label{fig:jhf-kt-bck1}
\end{figure}

We can now investigate the dependence of the $\sin^2 2\theta_{13}$ reach
on the detector parameters, and hence try to understand whether a
massive water Cherenkov detector is likely to be the best option.  In
Fig.~\ref{fig:jhf-kt-bck1} contours of constant $\sin^2 2\theta_{13}$
reach are shown as a function of the dataset size $D$ and the background
rate $f_B$ for 3 years of running at the 0.77~MW JHF beam (left-hand plots)
and at an upgraded 4~MW SJHF beam (right-hand plots).
The lower panels show how the $\sin^2 2\theta_{13}$ reach varies with
$\sigma_{f_B}/f_B$.
The contours have a characteristic shape. At sufficiently large $D$ the
$\sin^2 2\theta_{13}$ sensitivity is limited by the systematic
uncertainties associated with the background subtraction,
and the reach does not significantly improve with increasing dataset
size. The contours are therefore vertical in this region of
Fig.~\ref{fig:jhf-kt-bck1}. At sufficiently small $D$ the sensitivity
of the $\nu_\mu \rightarrow \nu_e$
appearance search is limited by signal statistics, and further
reductions in $f_B$ do not improve the $\sin^2 2\theta_{13}$ reach.
The contours are therefore horizontal in this
region of Fig.~\ref{fig:jhf-kt-bck1}.
The positions in the ($f_B$, $D$)-plane corresponding to our three
detector scenarios ($A$, $F$, and $W$) are indicated on the figure.
For the 0.77~MW machine the
two scenarios ($A$ and $W$) both yield reaches in the range $\sin^2
2\theta_{13} \sim$~0.015 to 0.03. However, the water Cherenkov
sensitivity is limited by the systematic uncertainty on the substantial
$\pi^0$ background. Hence, the $\sin^2 2\theta_{13}$ reach for scenario
$W$ does not improve substantially when the accelerator beam is upgraded
from 0.77~MW to 4~MW (SJHF). On the other hand this upgrade would result in a
substantial improvement in the reach obtained with scenario $A$, which
is not background limited, and therefore has a reach improving
almost linearly with $D$.  We conclude that, even with
SJHF, it will be difficult to observe a $\nu_\mu \rightarrow
\nu_e$ signal if $\sin^2 2\theta_{13}$ is less than about 0.01.
This conclusion is consistent with the JHF study group analysis, but is
in conflict with the expectations of Ref.~\cite{burt}.  On the positive
side, if $\sin^2 2\theta_{13}$ is larger than 0.01, a 1~GeV neutrino
beam at JHF or SJHF would permit the observation of a $\nu_\mu \rightarrow
\nu_e$ signal. Detector scenario $W$ does slightly better for a 0.77~MW
JHF, while scenario $A$ does slightly better for a 4.0~MW SJHF.

\begin{figure}
\centering\leavevmode
\epsfxsize=6.0in\epsffile{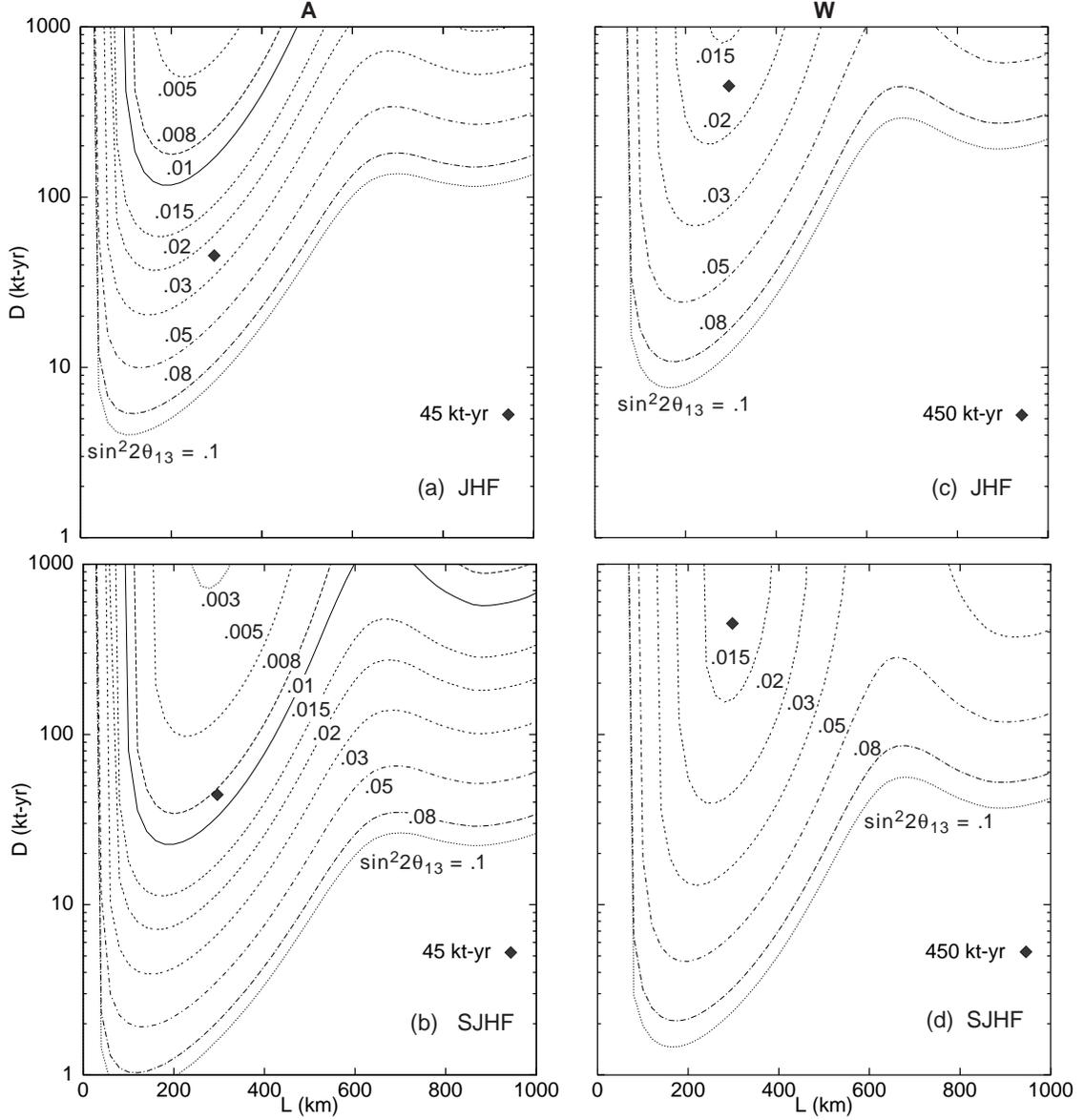}

\medskip
\vspace{0.5cm}
\caption[]{Contours of constant $\sin^22\theta_{13}$ reach that
correspond to a $\nu_e\to \nu_\mu$ signal that is 3 standard deviations
above the background. The contours are shown in the $(D, L)$-plane,
where $D$ is the data-sample size and $L$ the baseline.
The panels show predictions for
the JHF scenario described in the text, for detector
scenarios $A$ (left-hand plots) and $W$ (right-hand plots),
and for 0.77~MW (top plots) and 4.0 MW (bottom plots) proton drivers.
The positions corresponding to scenarios
$A$ and $W$ (see Table~\ref{tab:xx}) at $L = 295$~km are indicated.
}
\label{fig:jhf-kt-L}
\end{figure}

Finally, we consider whether the $\sin^2 2\theta_{13}$ reach at a
1~GeV JHF or SJHF neutrino beam can be improved with a different choice of
baseline. Contours of constant reach in the ($L, D$)-plane are
shown for scenarios $A$ and $W$ in Fig.~\ref{fig:jhf-kt-L}. A baseline
of 295~km does indeed yield the optimal reach for the water Cherenkov
scenario. For scenario $A$, a slightly shorter baseline (200~km) would
yield a slightly improved reach.

\subsection{Sensitivities for long baseline experiments}

\subsubsection{Decay channel length restrictions}

\begin{figure}[h]
\centering
\leavevmode
\epsfxsize=4.5in
\epsffile{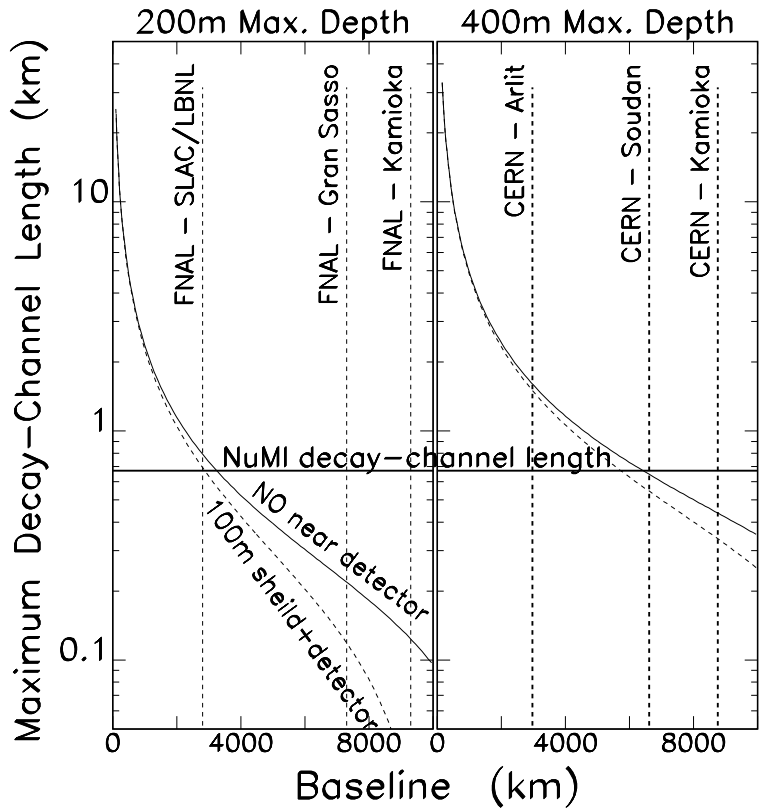}
\caption[]
{Maximum length of the pion decay channel that fits within a rock layer
that is 200~m deep (left-hand plot) and 400~m deep (right-hand plot)
shown as a function of baseline. The calculation is
described in the text. The solid (broken) curves shows the results without
(with) a near detector. For comparison, the horizontal solid line indicates
the NuMI decay channel length.}
\label{fig:channel_length}
\end{figure}

Consider next the sensitivity  that can be achieved with
longer baselines and higher energies.
We begin by considering how restrictions on the decay channel length
reduce the neutrino flux for very long baselines.
In a conventional neutrino beamline design it is desirable that the
pion decay channel is long enough for most of the pions to decay.
However, for very-long-baseline experiments the decay channel must
point downwards at a steep angle, and the geology under the
accelerator site may impose significant constraints on the maximum
length of the decay channel. In practice, an upgraded long-baseline
conventional neutrino beam would be sited at an existing particle
physics laboratory having a high-energy proton accelerator:
Brookhaven or Fermilab in the US, CERN or DESY in Europe, or the
planned JHF laboratory in Japan.
The rock characteristics under the JHF site are expected to be
determined next year by drilling~\cite{private_comm}.
The site with the deepest viable
rock layer in the US is Fermilab, which sits above approximately
200~m of good rock. The Brookhaven and DESY~\cite{norbert}
laboratories sit just above the
water table --- an impediment that would have to be overcome before
a high-energy long-baseline beam could be proposed.
The depth of the good rock (Molasse) under CERN
varies between about 200~m and 400~m, depending on location~\cite{mario}.
The impact of these restrictions on the maximum decay channel length
is shown as a function of the baseline length in
Fig.~\ref{fig:channel_length} for the
Fermilab and CERN sites. The resulting fraction of pions that decay
within the decay channel is summarized in Table~\ref{tab:x}
for several neutrino beam energies.
The channel length calculations were performed assuming that
(i)~the proton accelerator is at a depth of 10~m,
(ii)~the beam is then bent down to
point in the appropriate baseline-dependent direction using a magnetic
channel with an average field of 2~Tesla, and
(iii)~once pointing in the right
direction the proton beam enters a 50~m long targeting and focusing
section, after which the decay channel begins. The maximum decay channel
length then depends upon whether the channel extends all the way to the
bottom of the usable rock layer, or whether this rock layer must also
accommodate a near detector. Results for both of these cases
are presented in Fig.~\ref{fig:channel_length} and Table~\ref{tab:x}. In the
near-detector case the
maximum decay channel length has been reduced by 100~m to allow for
the shielding and detector hall. The pion decay fraction estimates
have been made assuming that all of the decaying pions have the
average pion energy in the channel.

The decay fractions in Table~\ref{tab:x} show that the
site-dependent depth
restrictions will result in a significant reduction in the neutrino
beam intensities for high-energy long-baseline beams. For example,
at the Fermilab site there is no room for a near detector if the
baseline is 9300~km (Fermilab to SuperK). With the medium energy beam
and a  baseline of 7300~km (Fermilab to Gran Sasso)
only 17\% of the pions decay within the channel. Hence, the
channel length restrictions would exclude, or at least
heavily penalize,  the extremely-long-baseline ideas proposed by
Dick et al.~\cite{freund-et-al}. Clearly, decay channel length
restrictions must be taken into account when comparing choices of
baseline and beam energy.

\subsubsection{$\sin^22\theta_{13}$ reach for scenarios $A$ and $W$}

We are now ready to consider the $\sin^2 2\theta_{13}$ reach that can be
obtained in a long-baseline experiment.
In our main discussion we consider
$\nu_e \to \nu_\mu$ appearance with $\delta m^2_{32} > 0$;
the $\delta m^2_{32} < 0$ case is
discussed at the end of this section. Our calculations use
the WBB spectra and interaction rates presented in the MINOS design
report~\cite{minos}
for the low-energy (LE) horn configuration ($E_\nu \sim 3$~GeV),
the medium-energy (ME) horn configuration ($E_\nu \sim 7$~GeV),
and the high-energy (HE) horn configuration ($E_\nu \sim 15$~GeV).
After accounting for the decay channel length restrictions arising
from a maximum depth requirement of 200~m, the neutrino
fluxes are assumed to scale with the inverse square of the baseline length.

The calculated reaches are listed in Tables~\ref{tab:y} and \ref{tab:z} for
detector scenarios $A$ and $W$, and several baselines: $L = 730$~km
(Fermilab $\rightarrow$ Soudan or CERN $\rightarrow$ Gran Sasso),
$L = 2900$~km (Fermilab $\rightarrow$ LBNL/SLAC),
$L = 7300$~km (Fermilab $\rightarrow$ Gran Sasso), and
$L = 9300$~km (Fermilab $\rightarrow$ SuperK).
Note that the shortest baseline
(730~km) has a very limited $\sin^2 2\theta_{13}$ reach
for all the beams, and the lowest energy beam (LE) has a very limited
$\sin^2 2\theta_{13}$ reach for all baselines. The best reach for
detector scenario $A$
is $\sin^2 2\theta_{13} = 0.003$, which
is obtained with a baseline that is
not too long (e.g.\ 2900~km). The best reach for detector scenario $W$
is also $\sin^2 2\theta_{13} = 0.003$, and is
obtained with long baselines (e.g.\ 7300~km or 9300~km) which benefit
from the enhancement of the oscillation amplitude due to
matter effects. The reaches for the two longest baselines are about
the same since the increase of the matter enhancement as $L$ increases
is compensated by the decrease in the pion decay fraction due to the
decay channel length restriction.

To further illustrate the impact of the decay channel length
restrictions
on the $\sin^2 2\theta_{13}$ reach for long-baseline experiments,
in Fig.~\ref{fig:kt-bck1}
contours of constant reach are shown in the ($f_B$, $D$)-plane
for $L = 7300$~km with the channel length
restrictions (right-hand plots) and
without the channel length restrictions (left-hand plots).
The scenario $W$ point lies in the
systematics-dominated (vertical contour) region, and is therefore not
significantly affected by a reduction in $D$ due to the decay channel
length restrictions.  However, the scenario $A$ point lies
between the systematics-limited and statistics-limited regions of the
plot, and
is significantly affected by the reduction in neutrino flux due to
the channel length restrictions.
Indeed, in scenario $A$ the $\sin^2 2\theta_{13}$ reach
at the HE beam is degraded from about 0.0015 to about 0.004 by the
channel length restriction.

\subsubsection{Dependence on detector parameters}

\begin{figure}
%\vspace{-0.2cm}
\centering\leavevmode
\epsfxsize=4.5in\epsffile{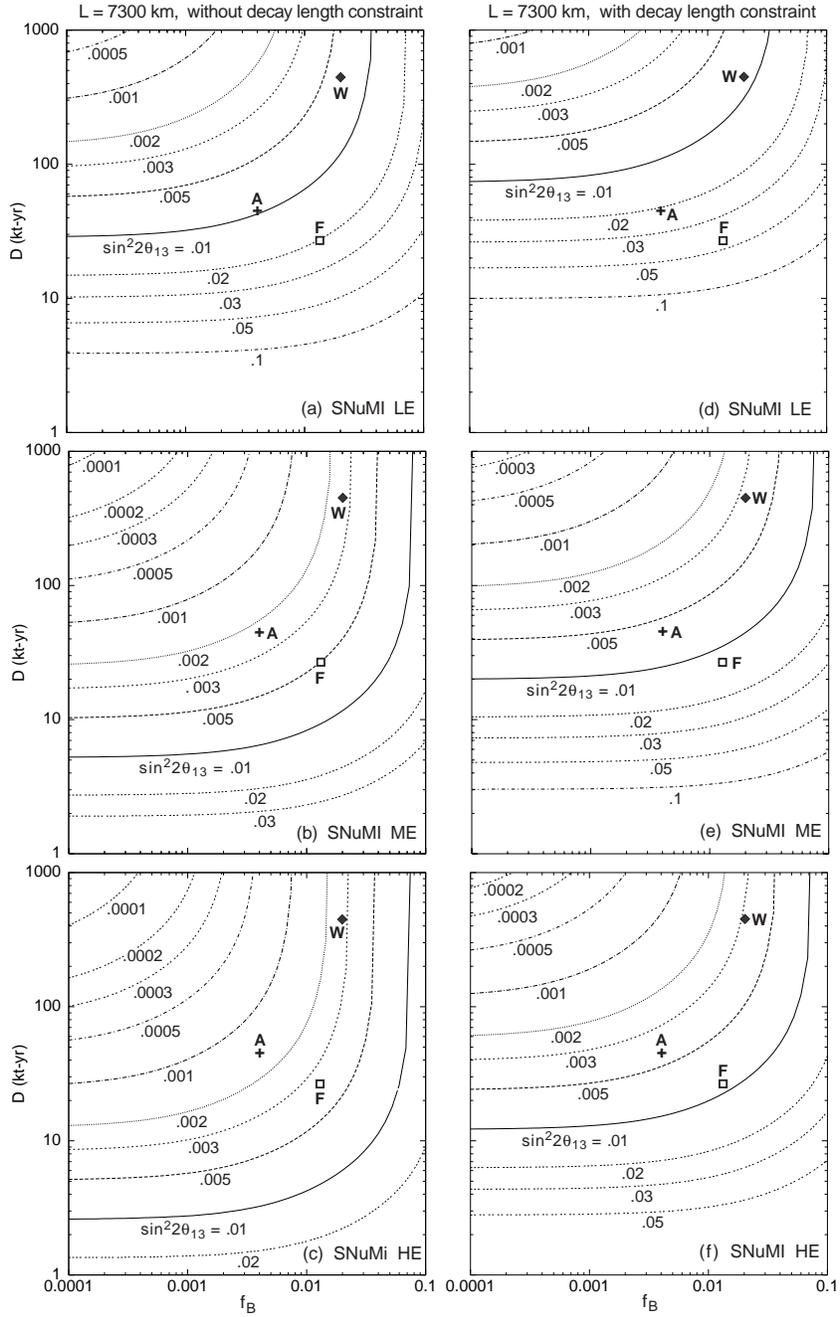}

\medskip
\vspace{0.5cm}
\caption[]{Contours of constant $\sin^22\theta_{13}$ reach that
correspond to a $\nu_e\to \nu_\mu$ signal that is 3 standard deviations
above the background, at $L = 7300$~km.
The contours are shown in the $(D, f_B)$-plane,
where $D$ is the data-sample size and $f_B$ the background rate divided
by the total CC rate.
The contours are shown for the LE (top plots), ME (center plots),
and HE (lower plots) upgraded SNuMI beams, both with (right plots)
and without (left plots) the decay length
constraint. The systematic uncertainty $\sigma_{f_B}/f_{B} = 0.1$.
The positions corresponding to the three standard
scenarios defined in Table~\ref{tab:xx} are indicated.}
\label{fig:kt-bck1}
\end{figure}

\begin{figure}
%\vspace{-1.0cm}
\centering\leavevmode
\epsfxsize=4.5in\epsffile{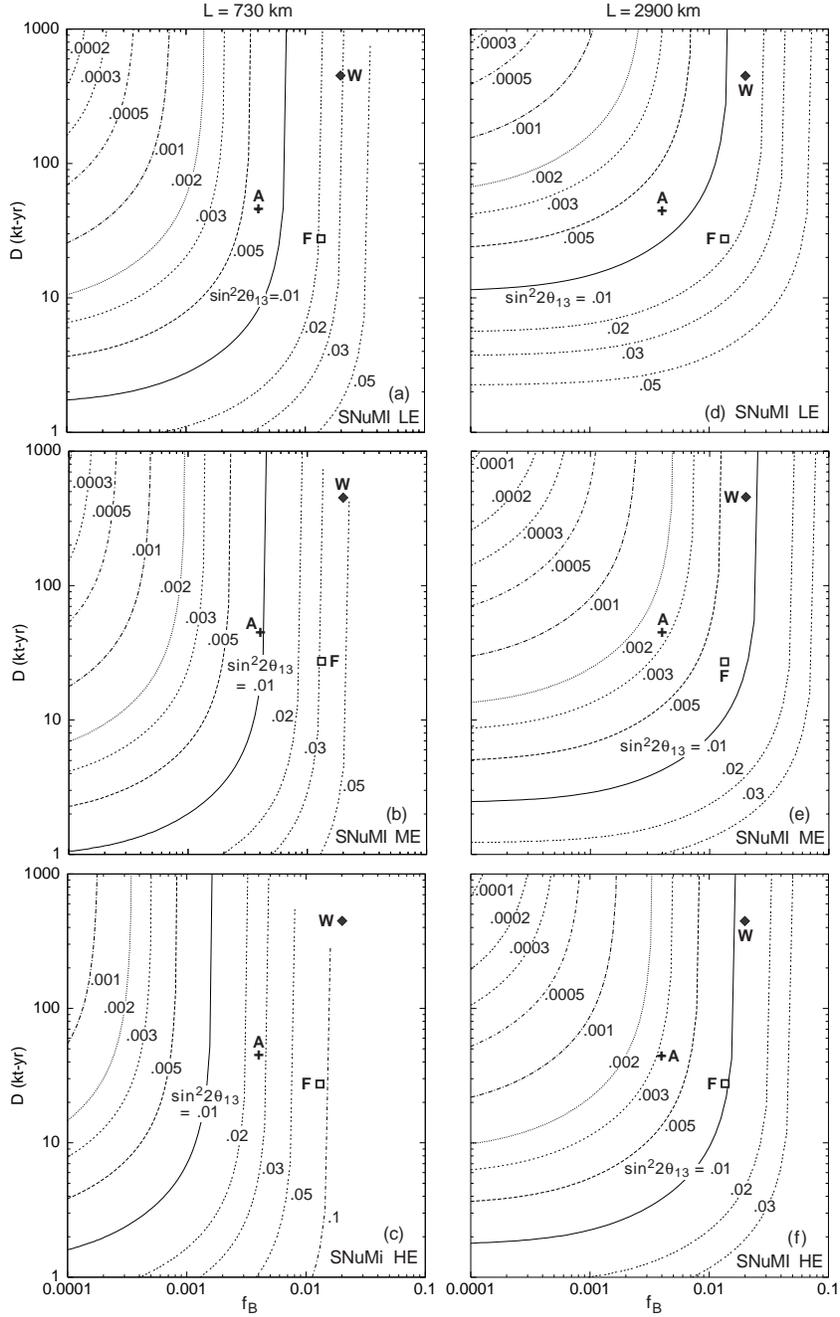}

\medskip
\vspace{0.5cm}
\caption[]{Contours of constant $\sin^22\theta_{13}$ reach that
correspond to a $\nu_e\to \nu_\mu$ signal that is 3 standard deviations
above the background, at $L = 730$~km (left plots) and 2900~km
(right plots).
The contours are shown in the $(D, f_B)$-plane,
where $D$ is the data-sample size and $f_B$ the background rate divided
by the total CC rate.
The contours are shown for the LE (top plots), ME (center plots),
and HE (lower plots) upgraded SNuMI beams.
The systematic uncertainty on the background subtraction is
$\sigma_{f_B}/f_{B} = 0.1$. The positions of
the three standard scenarios defined in
Table~\ref{tab:xx} are indicated.}
\label{fig:kt-bck2}
\end{figure}

\begin{figure}
\vspace{-0.5cm}
\centering\leavevmode
\epsfxsize=4.5in\epsffile{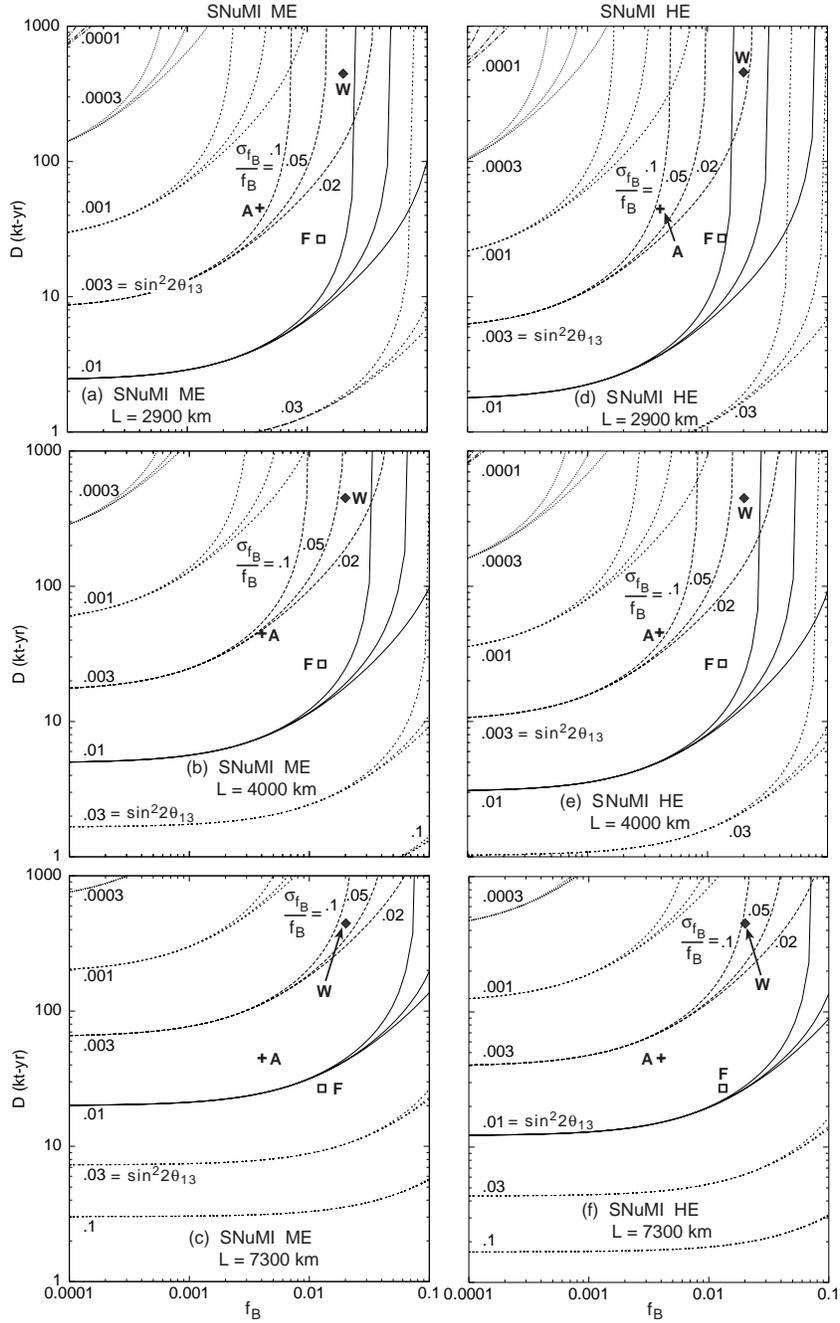}

\medskip
\vspace{0.5cm}
\caption[]{Contours of constant $\sin^22\theta_{13}$ reach that
correspond to a $\nu_e\to \nu_\mu$ signal that is 3 standard deviations
above the background, at the upgraded SNuMI ME (left) and HE
(right) beams.
The contours are shown in the $(D, f_B)$-plane,
where $D$ is the data-sample size and $f_B$ the background rate divided
by the total CC rate.
The contours are shown for
$L = 2900$ (top), $4000$ (center), and $7300$~km (bottom).
Curves are shown for
systematic uncertainties on the background subtraction
$\sigma_{f_B}/f_{B} = 0.1$, $0.05$, and
$0.02$. The positions
of the three standard scenarios defined in
Table~\ref{tab:xx} are indicated.
The decay length constraints have been imposed for
$L = 4000$ and $7300$~km.}
\label{fig:kt-bck3}
\end{figure}

We can now explore the dependence of the $\sin^2 2\theta_{13}$ reach on
the
baseline, beam energy, and detector parameters.
In Fig.~\ref{fig:kt-bck2}
contours of constant $\sin^2 2\theta_{13}$ reach are shown in
the ($f_B$, $D$)-plane for $L = 730$~km and 2900~km.
As already noted, for our
detector scenarios $A$ and $W$ the best reach is $\sim 0.003$, obtained
with scenario $A$ at $L = 2900$~km using either the ME or HE beams
(Figs.~\ref{fig:kt-bck2}e and \ref{fig:kt-bck2}f), or with scenario $W$
at $L = 7300$~km using the same beams (Figs.~\ref{fig:kt-bck1}e and
\ref{fig:kt-bck1}f).
It is interesting to consider what improvements to scenarios $A$ and
$W$ would be required to obtain a reach of 0.001, for example.
This goal can be attained by decreasing the background
fraction $f_B$ for scenario $W$ to $f_B \sim 0.004$
(or alternatively increasing the dataset size $D$ for scenario $A$ by
a factor of 10) and using the HE beam at a very long baseline.
The goal could also be attained by decreasing $f_B$ for scenario
$A$ by an order of magnitude and using the high energy beam and a baseline
of 2900~km, for example. None of these revised detector scenarios
seems practical. An alternative strategy is to try to find a detector
scenario with a smaller systematic uncertainty on $f_B$.
Fig.~\ref{fig:kt-bck3} shows, for the ME and HE beams, contours of constant
$\sin^2 2\theta_{13}$ reach in
the ($f_B$, $D$)-plane for several different $\sigma_{f_B}/f_B$, and
for baselines of 2900~km, 4000~km, and 7300~km.
The scenario $W$ sensitivity would benefit if the systematic uncertainty
on the background could be reduced, but even a factor of five improvement
in $\sigma_{f_B}/f_B$ would not
permit a reach of 0.001 to be attained.

Since detector scenarios $A$ and $W$ are ambitious, we can ask what
happens
if $D$, $f_B$, or $\sigma_{f_B}/f_B$ must be relaxed.
The best reaches obtained with scenario $W$ were for the
ME and HE beams at very long baselines (e.g. $L =7300$~km).
In these cases the reach is not very sensitive to $D$, but degrades
roughly
linearly with increasing $f_B$ (Fig.~\ref{fig:kt-bck1}) or
$\sigma_{f_B}/f_B$
(Fig.~\ref{fig:kt-bck3}). Hence, if the achievable
background rate is really $f_B = 0.1$ then the $\sin^2 2\theta_{13}$
reach
is well above 0.01 for the observation of a $\nu_\mu \rightarrow \nu_e$
signal at
3 standard deviations above the background.
The best reaches obtained with scenario $A$ were for the
ME and HE beams at long baselines (e.g. $L =2900$~km).
In these cases the reach is sensitive to both decreases in $D$ and
increases in $f_B$.
The reach can be degraded by a factor of 2 by either reducing $D$ by
about a factor
of 4, or by increasing $f_B$ by about a factor of 3 (Fig.~\ref{fig:kt-bck2}).

So far we have considered only a few discrete baseline lengths.
To explore the reach that can be obtained with other baseline choices,
Fig.~\ref{fig:kt-L}
shows, for each of the three NuMI beam energies,
contours of constant $\sin^22\theta_{13}$ reach in the ($L$, $D$)-plane
for scenarios $A$ and $W$.
For scenario $A$, where backgrounds are less important, the
optimal distance varies with beam energy; crudely speaking, the optimal
$L$ is given by making the vacuum oscillation argument $1.27 \delta
m^2_{32} L/\langle E_\nu \rangle$ of order $\pi/2$. For scenario $W$ the
backgrounds are more important and larger distances give a better
$\sin^22\theta_{13}$ reach for all three upgraded NuMI beams.

Finally, we have also studied neutrino beams with higher energy
than NuMI. For example, the CNGS beam~\cite{cngs} at CERN has an average
neutrino energy of about 20~GeV. We find that for the expected
$3\times10^{19}$ protons on target per year, three years of running will
at best give a $\sin^22\theta_{13}$ reach of about 0.01 for either
scenario $A$ or $W$. Upgrading the
proton intensity by a factor of four improves the $\sin^22\theta_{13}$
reach to about 0.005. Therefore we conclude that the higher-energy
CNGS superbeams have similar capability to the SNuMI beams.

\subsubsection{Summary of $\sin^22\theta_{13}$ reaches for
$\delta m^2_{32} > 0$ and $\delta m^2_{32} < 0$}

\begin{figure}
%\vspace{-1.0cm}
\centering\leavevmode
\epsfxsize=4.5in\epsffile{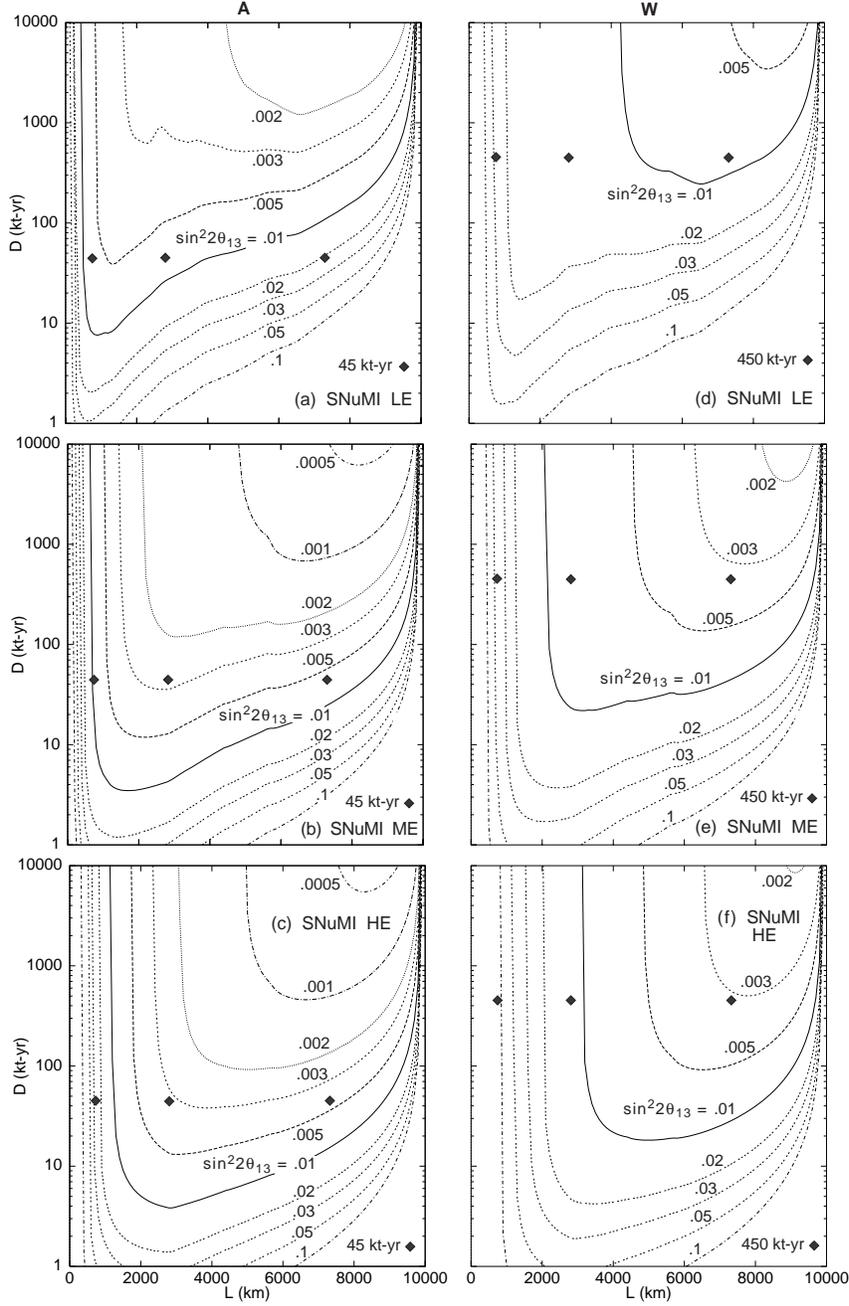}

\medskip
\vspace{0.5cm}
\caption[]{Contours of constant $\sin^22\theta_{13}$ reach that
correspond to a $\nu_e\to \nu_\mu$ signal that is 3 standard deviations
above the background. The contours are shown in the $(D, L)$-plane,
where $D$ is the data-sample size and $L$ the baseline.
The panels show predictions for
the upgraded SNuMI LE (top), ME (center), and HE (bottom) beams,
and for detector
scenarios $A$ (left plots) and $W$ (right plots).
The decay length constraint is included.
The systematic uncertainty on the background subtraction
$\sigma_{f_B}/f_{B} = 0.1$. The positions of the
standard scenarios defined in Table~\ref{tab:xx} are shown
at $L = 730$, $2900$, and $7300$~km.}
\label{fig:kt-L}
\end{figure}

In summary, Figs.~\ref{fig:kt-bck1}, \ref{fig:kt-bck2}, and
\ref{fig:kt-L} show that the best $\sin^22\theta_{13}$
reach that can be obtained with detector scenarios $A$ and $W$
is about 0.003.
This optimum reach can be obtained in scenario $A$ with $L \sim
2000$--$4000$~km for the NuMI ME beam or with $L \sim 3000$--$6000$~km
for the HE beam, or in scenario $W$ with $L \sim 7000$--$9000$~km for
either the ME or HE beams. Scenarios $A$ and $W$ require ambitious
detector parameters. To improve the reach to 0.001, for example, requires
substantial improvements in $f_B$, $\sigma_{f_B}/B$, and/or $D$, and
does not therefore seem practical. If the scenario $A$ and $W$
parameters cannot be
realized the reach will be degraded. In particular, a significant
increase of $f_B$ (or $\sigma_{f_B}/f_B$) in either scenario $A$ or $W$ would
result in a significant decrease in $\sin^22\theta_{13}$ reach. A
significant decrease in the data-sample size in scenario $A$ will also
degrade the $\sin^22\theta_{13}$ reach.

Up to now we have considered the sensitivity of long baseline
experiments
if $\delta m^2_{32} > 0$. We now turn our attention to the alternative
case: $\delta m^2_{32} < 0$. In this case
long baseline experiments using a neutrino beam will suffer
from a suppression of signal due to matter effects.
Therefore, in our scenarios $A$ and $W$, if no signal is observed after
3 years of neutrino running the beam is switched to antineutrinos for a
further 6 years of data taking.
For antineutrino running with $\delta m^2_{32} < 0$ the results shown
in Figs.~\ref{fig:kt-bck1}, \ref{fig:kt-bck2}, and
\ref{fig:kt-L} must be modified
since the antineutrino cross section is about half of the
neutrino cross section. Hence we must double the required values on the
$D$-axes in the various figures. Other modifications to the contour
plots
for antineutrino running with $\delta m^2_{32} < 0$ are minor
since the matter enhancement in this case is similar to
the enhancement for neutrinos when $\delta m^2_{32} > 0$ (they are the
same in the limit that the sub-leading oscillation can be ignored).
However, the positions of the scenario $A$ and $W$ points on the various
figures must be moved to account for the larger values of $D$ and
$f_B$ (and potentially $\sigma_{f_B}/f_B$).
Note that the larger background rate associated with antineutrino
running in Scenario $A$ will degrade the ultimate $\sin^22\theta_{13}$
reach for $\delta m^2_{32} < 0$; the best reach becomes $\sim 0.004$.
%Looking for $\nu_e \to \nu_\mu$ when
%$\delta m^2_{32} < 0$ (or $\bar\nu_e \to \bar\nu_\mu$ when $\delta
%m^2_{32} > 0$) will be difficult, since in each case the effects of
%matter suppress the oscillation. In these cases, an appearance signal is
%best seen in the channel with opposite $CP$.

\section{Neutrino mass hierarchy and CP-violation}
\label{sec:cpv}

In the 1~GeV and multi-GeV superbeam scenarios that we have
considered it will be difficult to observe a $\nu_\mu \rightarrow
\nu_e$
or $\bar\nu_\mu \to \bar\nu_e$ signal if $\sin^2 2\theta_{13}$ is
smaller than about 0.01. However, if $\sin^2 2\theta_{13}$ is
$\cal O$(0.01) a
$\nu_\mu \rightarrow \nu_e$ signal would be observable
provided a sufficiently massive
detector with sufficiently small background is practical.
We would like to know if, in this case, the sign of $\delta m^2_{32}$
can be determined in the long-baseline multi-GeV beam experiment,
and whether $CP$ violation might be observed in either the
long-baseline multi-GeV beam experiment or the 1~GeV intermediate
baseline experiment. We begin by considering the $CP$ sensitivity
at the SJHF, and then consider the sensitivity for determining
$CP$ violation and/or the pattern of neutrino masses at long baselines.

\subsection{$CP$ violation with a JHF superbeam}

\begin{figure}
\centering\leavevmode
\epsfxsize=6.0in\epsffile{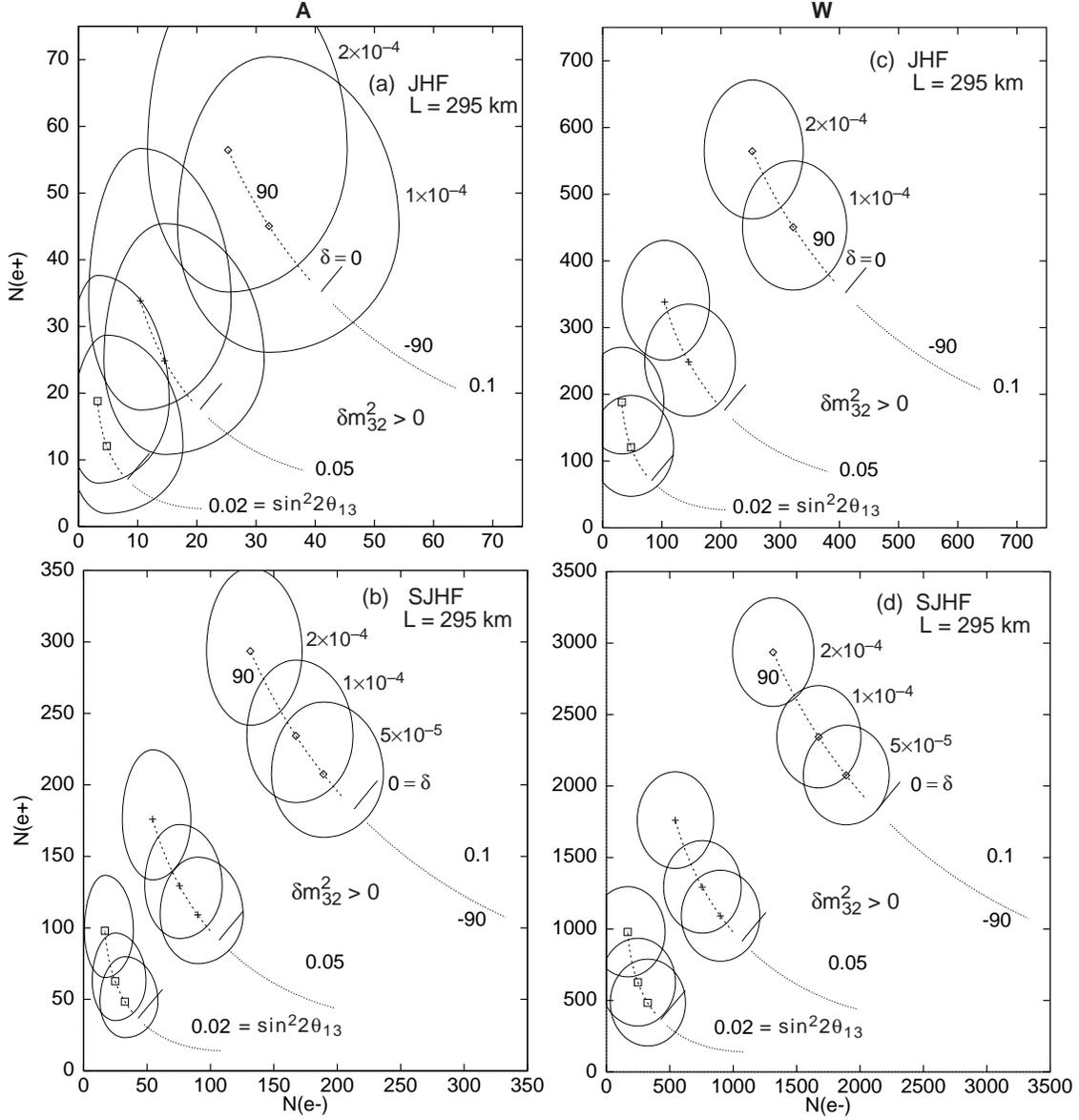}

\medskip
\vspace{0.5cm}
\caption[]{$3\sigma$ error ellipses in the
$\left[N(e^+), N(e^-)\right]$-plane, shown for the 0.77~MW JHF (top
plots) and 4~MW SJHF(bottom plots) scenarios with
at $L = 295$~km. The contours are shown for detector scenarios
$A$ (left) and $W$ (right), with
$\sin^22\theta_{13} =
0.02$, $0.05$, and $0.1$. The solid (dashed) [dotted] curves correspond
to $\delta = 0^\circ$ ($90^\circ$) [$-90^\circ$] with $\delta m^2_{21}$
varying from $2\times10^{-5}$~eV$^2$ to $2\times10^{-4}$~eV$^2$.
The error ellipses are shown for three simulated data points
at $\delta m^2_{21} = 5\times10^{-5}$, $10^{-4}$ and
$2\times10^{-4}$~eV$^2$.}
\label{fig:jhfcpv}
\end{figure}

In our SJHF scenario, a $CP$ violation search would consist of running for
3~years with a neutrino beam and measuring the number of $\nu_\mu
\rightarrow \nu_e$ signal events [$N(e-)$], and then running for 6~years
with an antineutrino beam and measuring the number of $\bar\nu_\mu \to
\bar\nu_e$ signal events [$N(e+)$].  In our calculations we assume that
the antineutrino cross section is about one-half of the neutrino cross
section, and that the antineutrino flux is the same as the neutrino
flux. In the absence of $CP$ violation ($\delta = 0$ or $180^\circ$),
after correcting for cross-section and flux differences, we would
therefore expect $N(e+) \simeq N(e-)$. In the presence of
maximal $CP$ violation with $\delta = 90^\circ [-90^\circ]$ we would
expect $N(e+) > N(e-)$ $[N(e+) < N(e-)]$. The magnitude of the deviation
from $N(e+) = N(e-)$ induced by $CP$ violation is quite sensitive to the
sub-leading scale $\delta m^2_{21}$. Setting $\delta =90^\circ$,
in Fig.~\ref{fig:jhfcpv} the
predicted positions in the [$N(e-), N(e+)$]-plane are shown for
scenarios $A$ (left-hand plots) and $W$ (right-hand plots)
The predictions are shown as a function of both
$\sin^2 2\theta_{13}$ and $\delta m^2_{21}$.
The error ellipses around each point indicate the measurement
precision at 3~standard deviations, taking into account both statistical
and systematic uncertainties, and using the statistical prescription
described in the appendix. An overall normalization uncertainty (which
could account for uncertainties in the flux and/or cross sections) of
2\% is included, although its effects are generally small.
$CP$ violation can be established at the
$3\sigma$ level if the error ellipses do not overlap the
$CP$-conserving curves (solid lines, $\delta = 0$). The curves for the
other $CP$-conserving case ($\delta = 180^\circ$) lie very close to the
$\delta = 0$ curves and are not shown.

Note that
for scenario $W$ with the upgraded 4~MW SJHF beam,
if $\sin^2 2\theta_{13} = 0.1$
(larger values are already excluded), $\delta m^2_{21} = 5 \times
10^{-5}$~eV$^2$, and $\delta = 90^\circ$, then the predicted point in
the [$N(e-), N(e+)$]-plane is just $3\sigma$ away from the
$CP$ conserving ($N(e+) = N(e-)$) prediction.  Alternatively, if $\sin^2
2\theta_{13} = 0.02$, $\delta m^2_{21} = 1 \times 10^{-4}$~eV$^2$
(larger values are improbable), and $\delta = 90^\circ$, then the
predicted point is also just $3\sigma$ away from the $CP$-conserving
prediction.  Hence there is a small region of the allowed parameter
space ($\sin^2 2\theta_{13} > 0.02$ and $\delta m^2_{21} > 5 \times
10^{-5}$~eV$^2$) within which maximal $CP$ violation might be observable
at an upgraded JHF if $\delta m^2_{32}$ is in the center of the presently
favored SuperK region and $\sin^2 2\theta_{23} \sim 1$. It is also
possible to detect maximal $CP$ violation for $\sin^2 2\theta_{13} >
0.05$ and $\delta m^2_{21} > 10^{-4}$~eV$^2$ with the 0.77~MW JHF in the
$W$ scenario. Generally
detector scenario $W$ does better for $CP$ violation, except scenario $A$
is slightly better for $\sin^22\theta_{13} \simeq 0.02$ at the 4~MW SJHF.
Because the matter effect is small at $L = 295$~km, predictions
for $\delta m^2_{32} > 0$ and $\delta m^2_{32} < 0$ are nearly the same,
and hence the sign of $\delta m^2_{32}$ cannot be determined.

\subsection{$CP$ violation and the sign of $\delta m^2_{32}$
at long-baseline experiments}

\begin{figure}
\centering\leavevmode
\epsfxsize=6.0in\epsffile{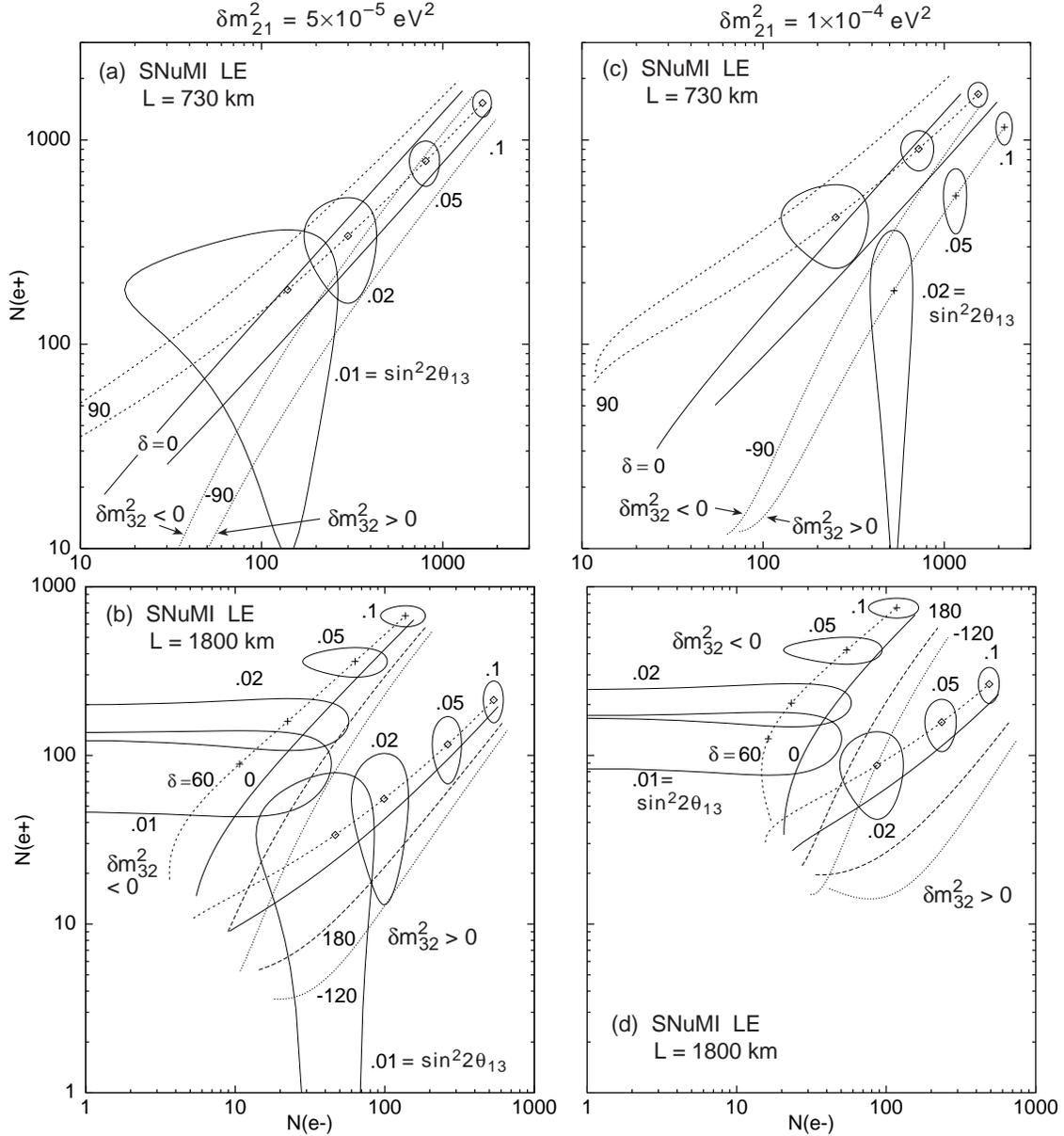}

\medskip
\vspace{0.5cm}
\caption[]{$3\sigma$ error ellipses in the
$\left[ N(e^+), N(e^-) \right]$-plane, shown for
detector scenario $A$ at the
upgraded LE SNuMI beam with $L = 730$ (top
plots) and 1800~km (bottom plots).
The contours are shown for $\delta m^2_{21} = 5\times10^{-5}$ (left)
and $10^{-4}$~eV$^2$ (right).
The solid and long-dashed curves
correspond to the $CP$ conserving cases $\delta = 0^\circ$ and
$180^\circ$, and the short-dashed and dotted curves correspond to two
other cases that give the largest deviation from the $CP$ conserving
curves; along these curves $\sin^22\theta_{13}$ varies from 0.001 to
0.1, as indicated.}
\label{fig:numicpv1}
\end{figure}

Consider next long-baseline experiments using multi-GeV neutrino beams.
The approximate equality $N(e+) \simeq N(e-)$ will be modified by
intrinsic $CP$ violation and by matter effects.  Predictions in the
[$N(e-), N(e+)$]-plane are shown in Fig.~\ref{fig:numicpv1} for
scenario $A$ using the 1.6~MW LE superbeam for two values of $\delta
m^2_{21}$ ($5 \times 10^{-5}$~eV$^2$ and $1 \times 10^{-4}$~eV$^2$), and
for two baselines ($L = 730$~km and 1800~km). The predictions for each
of these cases are shown as a function of $\sin^2 2\theta_{13}$,
$\delta$, and the sign of $\delta m^2_{32}$, with $|\delta m^2_{32}| =
3.5 \times 10^{-3}$~eV$^2$ and $\sin^2 2\theta_{23} = 1$.  Note that at
$L = 730$~km the magnitude of the modifications of the appearance rates
due to matter effects are comparable to the magnitudes of the
modifications due to maximal intrinsic $CP$ violation. Furthermore, the
expected precisions of the measurements, shown on the figure by the
$3\sigma$ error ellipses, are also comparable to the sizes of the
predicted $CP$ and matter effects.

\begin{figure}
%\vspace{-1.0cm}
\centering\leavevmode
\epsfxsize=4.5in\epsffile{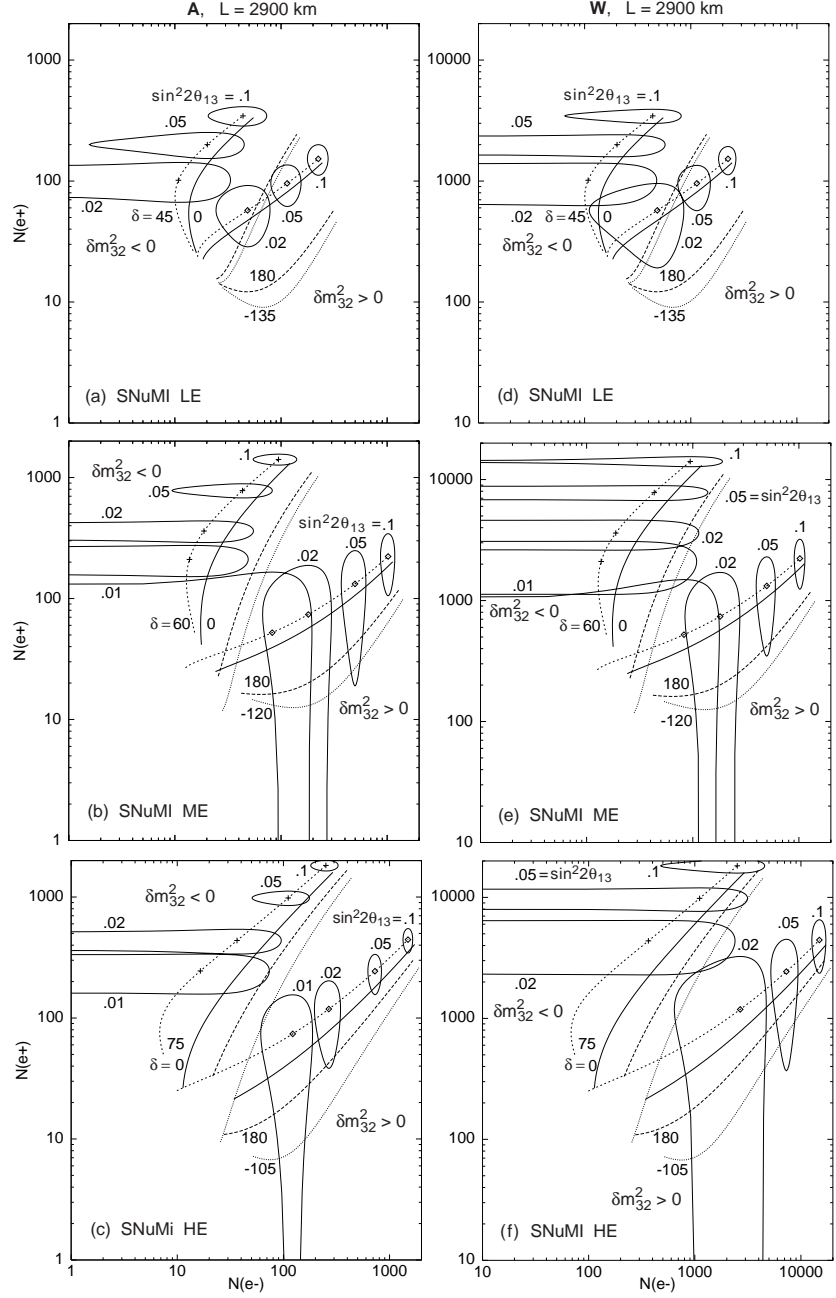}

\medskip
\vspace{0.5cm}
\caption[]{$3\sigma$ error ellipses in the
$\left[N(e^+), N(e^-)\right]$-plane, shown for
detector scenarios $A$ (left) and $W$ (right) at
$L = 2900$~km with the
upgraded LE (top), ME (center), and HE (bottom) SNuMI beams.
The contours are shown for $\delta
m^2_{21} = 10^{-4}$~eV$^2$. The solid and long-dashed curves
correspond to the $CP$ conserving cases $\delta = 0^\circ$ and
$180^\circ$, and the short-dashed and dotted curves correspond to two
other cases that give the largest deviation from the $CP$ conserving
curves; along these curves $\sin^22\theta_{13}$ varies from 0.001 to
0.1, as indicated.}
\label{fig:numicpv2}
\end{figure}

Matter effects will cause the two $CP$-conserving cases $\delta = 0$
and $\delta = 180^\circ$ to give different predictions for $N(e^+)$ and
$N(e^-)$, and therefore to establish $CP$ violation the signal must be
distinguishable from both $\delta = 0$ and $\delta = 180^\circ$. Hence,
in the scenario we are considering, Fig.~\ref{fig:numicpv1} shows that
superbeam measurements with the LE beam at 730~km can help to constrain
the parameter space, but generally cannot provide unambiguous
evidence for intrinsic $CP$ violation, and cannot unambiguously determine
the sign of $\delta m^2_{32}$. The only exception to this is if $\delta
m^2_{32} > 0$ and $\delta = - 90^\circ$ (or $\delta m^2_{32} < 0$ and
$\delta = 90^\circ$), in which case $CP$ violation could
be established and the sign of $\delta m^2_{32}$ determined for
$\sin^22\theta_{13} > 0.02$. The $CP$ and matter effects are better
separated at $L = 1800$~km, for which an unambiguous determination of
the sign of $\delta m^2_{32}$ seems possible provided $\sin^2
2\theta_{13} > 0.02$, although $CP$ violation cannot be established for
$\delta m^2_{21} < 10^{-4}$~eV$^2$. At smaller values of $\sin^2
2\theta_{13}$ modifications to the appearance rates cannot distinguish
between matter and $CP$ effects. Note that, because of the matter
effect, at distances longer than 1000~km the values of $\delta$ that
give the largest disparity of $N(e^+)$ and $N(e^-)$ are no longer
$\pm 90^\circ$. Also note that the sign of $\delta m^2_{32}$ is most easily
determined when the $CPV$ and matter effects interfere constructively to
give a greater disparity of $N(e^+)$ and $N(e^-)$, and more difficult
when the $CPV$ and matter effects interfere destructively [i.e.,
$N(e^+)$ and $N(e^-)$ are more equal].
Going to even longer baselines, predictions in the
[$N(e-), N(e+)$] plane are shown in Fig.~\ref{fig:numicpv2} for scenario
$A$ (left-hand plots) and scenario $W$ (right-hand plots) with $L =
2900$~km. The predictions are shown for the LE beam (top plots), ME beam
(middle plots), and HE beam (bottom plots). In general, the sign of
$\delta m^2_{32}$ can be determined provided $\sin^2 2\theta_{13} >
0.02$, but in none of the explored long-baseline scenarios can
$CP$-violation be unambiguously established for $\delta m^2_{21} <
10^{-4}$~eV$^2$.

\subsection{CP-violation and the sign of $\delta m^2_{32}$
at a neutrino factory}

We can ask, how do the $CPV$ and $\delta m^2_{32}$-sign capabilities
of superbeams compare with those of a neutrino factory?  The relevant
experimental signature at neutrino factory is the appearance of a
wrong-sign muon indicating $\nu_e \rightarrow \nu_\mu$ (or
$\bar{\nu}_e \rightarrow \bar{\nu}_\mu$) transitions.  This is
a much cleaner signature than electron appearance with a superbeam.
Hence, background systematics are under better control at a neutrino
factory, and the expected error ellipses in the
[N($\mu+$), N($\mu-$)]-plane are therefore much smaller.

In our analysis we assume a 20~GeV neutrino factory with
$1.8\times10^{21}$ useful $\mu^+$ decays (which might be achieved in
three years running at at high-performance neutrino factory) and
$3.6\times10^{21}$ useful $\mu^-$ decays, a 50~kt iron--scintillator
detector~\cite{albright} at distances $L = 1800$~km, 2900~km, and
4000~km from the source. For comparison, the total neutrino flux for
three years running at a distance of 1~km from the source is
$2\times10^{19}/{\rm m}^2$ for the neutrino factory scenario, while it
is $7\times10^{16}/{\rm m}^2$ for SJHF and $4\times10^{18}/{\rm m}^2$
for the SNuMI HE beam. We choose an iron--scintillator detector for the
neutrino factory analysis since it is particularly well--suited for the
detection of muons and can be made larger than, e.g., a liquid argon
detector, at a similar or lower cost. We also take $f_B =10^{-4}$,
$\sigma_{f_B}/f_B = 0.1$, and a normalization uncertainty of 2\%. This
background level can be achieved with a 4~GeV cut on the detected muon,
which gives a detection efficiency of about 73\%, implying an effective
data sample of $D = 110$~kt--yr for three years running.

\begin{figure}
%\vspace{-1.0cm}
\centering\leavevmode
\epsfxsize=4.5in\epsffile{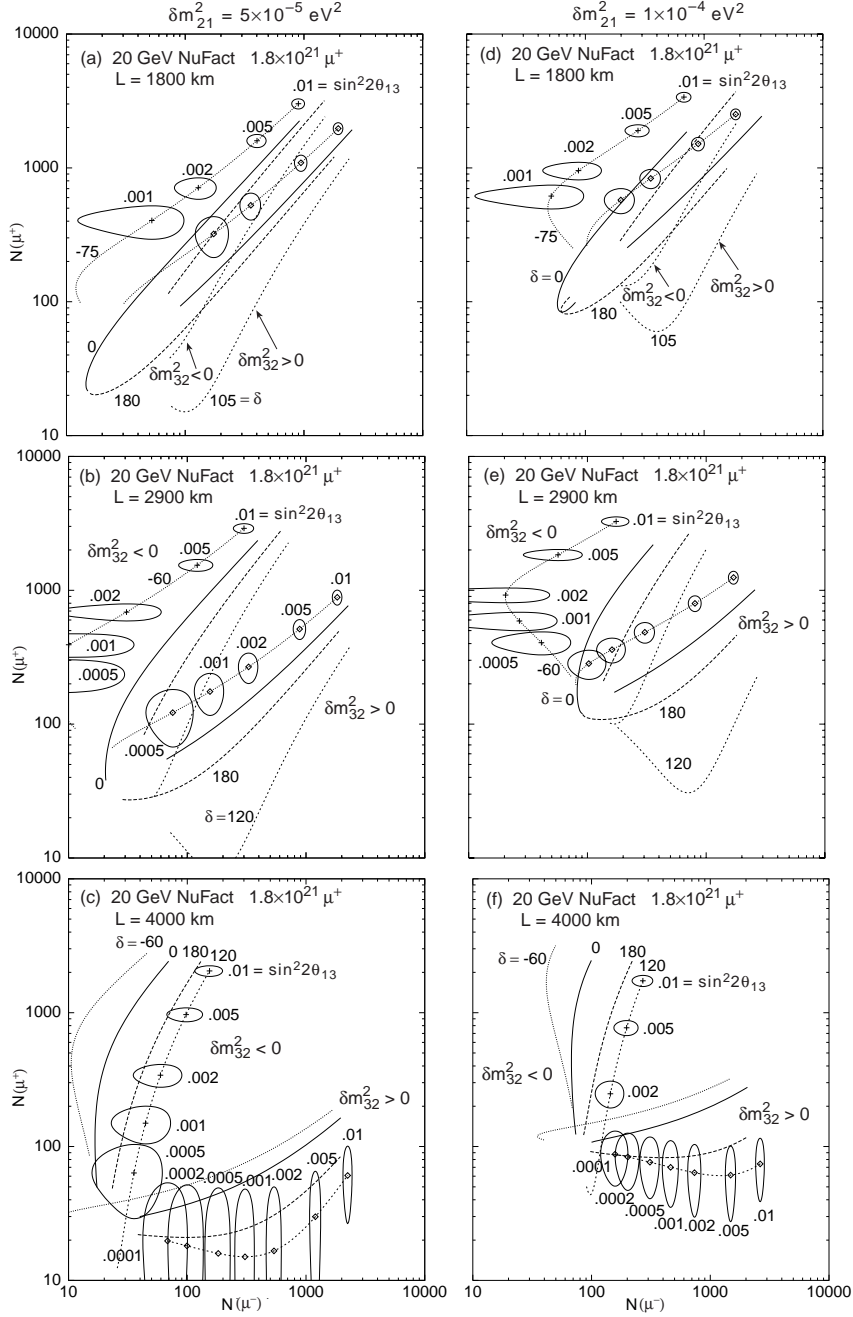}

\medskip
\vspace{0.5cm}
\caption[]{$3\sigma$ error ellipses in the
$\left[N(\mu^+), N(\mu^-)\right]$-plane, shown for a neutrino
factory delivering $3.6\times10^{21}$ useful decays of 20~GeV muons and
$1.8\times10^{21}$ useful decays of 20~GeV antimuons, with a 50~kt detector
at $L = 1800$ (top), $2900$ (center), and $4000$~km (bottom), with
$\delta m^2_{21} = 5\times10^{-5}$ (left) and $10^{-4}$~eV$^2$ (right).
The solid and long-dashed curves correspond to the $CP$ conserving cases $\delta =
0^\circ$ and $180^\circ$, and the short-dashed and dotted curves
correspond to two other cases that give the largest deviation from the
$CP$ conserving curves; along these curves $\sin^22\theta_{13}$ varies
from 0.0001 to 0.01, as indicated.}
\label{fig:nufactcpv}
\end{figure}

\begin{figure}
\centering\leavevmode
\epsfxsize=4.5in\epsffile{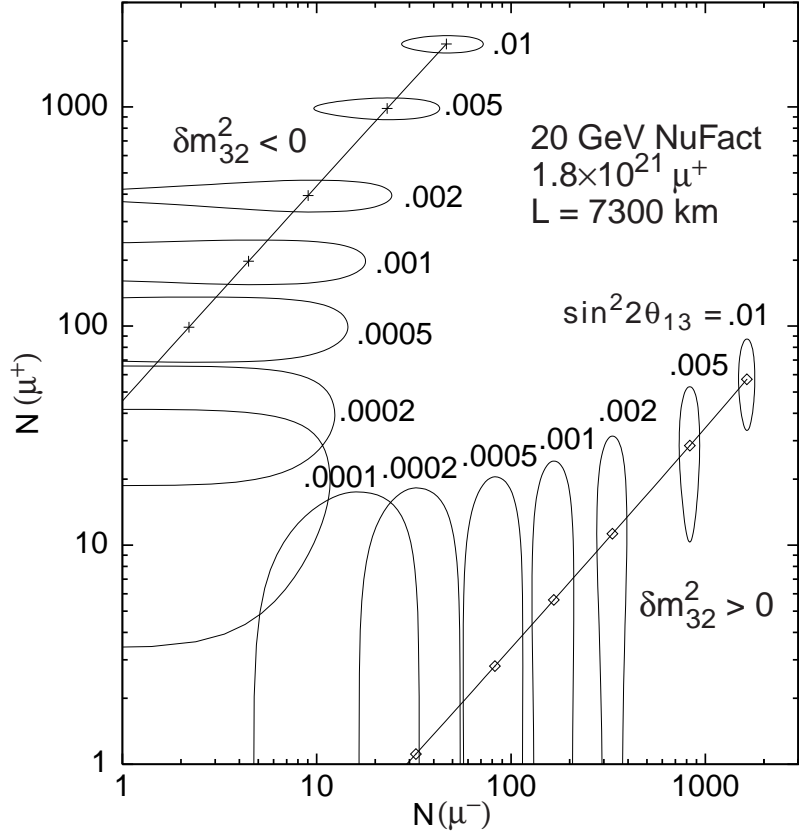}

\medskip
\vspace{0.5cm}
\caption[]{$3\sigma$ error ellipses in the
$\left[N(\mu^+), N(\mu^-]\right)$-plane, shown for a neutrino
factory delivering $3.6\times10^{21}$ useful decays of 20~GeV muons and
$1.8\times10^{21}$ useful decays of 20~GeV antimuons, with a 50~kt
detector at $L = 7300$~km, $\delta m^2_{21} = 10^{-4}$~eV$^2$,
and $\delta = 0$. Curves are shown for both signs of
$\delta m^2_{32}$; $\sin^22\theta_{13}$ varies along the curves from
0.0001 to 0.01, as indicated.
}
\label{fig:nufactcpv2}
\end{figure}

The corresponding neutrino factory predictions in the [N($\mu-$),
N($\mu+$)]-plane are shown in Fig.~\ref{fig:nufactcpv}. The 1800~km
baseline is too short, since matter and $CP$ effects are
indistinguishable in most cases. At 2900~km the predictions allow an
unambiguous determination of the sign of $\delta
m^2_{32}$ for much of the parameter space, and the possibility of
establishing the existence of $CPV$.
At 4000~km the statistical uncertainties are larger, and impair the
sensitivity to observe $CPV$.
However, matter effects are also larger, and an unambiguous determination
of the sign of $\delta m^2_{32}$ is possible down to $\sin^2 2\theta_{13}$
of a few $\times 10^{-4}$. For very long baselines (e.g. $L = 7300$~km,
Fig.~\ref{fig:nufactcpv2}) there is negligible sensitivity
to $CPV$ or to $\delta m^2_{21}$, matter effects are large,
and the $\sin^2 2\theta_{13}$ reach for
determining the sign of $\delta m^2_{32}$ approaches $10^{-4}$.

\section{Summary}

We have explored the oscillation-physics capabilities
of 1~GeV and multi-GeV neutrino beams produced at
MW-scale proton accelerator facilities (neutrino superbeams).
Specifically, the limiting value of $\sin^2 2\theta_{13}$
that would permit the first observation of
$\nu_\mu \rightarrow \nu_e$ and/or
$\bar{\nu}_\mu \rightarrow \bar{\nu}_e$ oscillations
at 3 standard deviations is considered, along with
the ability of these intense conventional neutrino beams
to determine the pattern of neutrino masses (sign of $\delta m^2_{32}$)
and discover $CP$-violation in the lepton sector.
The figures in this paper provide a toolkit for accessing the
physics capabilities as a function of the detector specifications,
characterized by the dataset size $D$~(kt-years) and the uncertainty
on the background subtraction (given by $f_B$ and $\sigma_{f_B}/f_B$).
Table~\ref{tab:s} summarizes the physics capabilities of some
beam-detector combinations. Also shown in the table are similar results
for an entry-level and high-performance neutrino factory with $E_\mu =
20$~GeV.

Determining the optimum detector technology and characteristics is
beyond the scope of this paper, and may require a detector R\&D program.
However, for some ambitious but plausible detector scenarios we find:
\begin{description}
\item{(i)} With a sufficiently ambitious detector, if
$\sin^2 2\theta_{13} >$ few $\times 10^{-3}$ and $\delta m^2_{32} > 0$,
then $\nu_\mu \rightarrow \nu_e$ and $\bar{\nu}_\mu \rightarrow
\bar{\nu}_e$ signals should be observable at a superbeam. The
reach is slightly worse if $\delta m^2_{32} < 0$. The best reach is
obtained with a long-baseline multi-GeV superbeam; for example, with the
SNuMI ME or HE beams and a baseline $\ge 2000$~km. This would permit the
tightening of constraints on the oscillation parameter space. It is
important to account for decay channel length restrictions when
assessing the capabilities of very-long-baseline experiments.
\item{(ii)} If $CP$ is maximally violated in the lepton sector,
there is a small region of allowed parameter space in which an
experiment at a JHF or SJHF beam ($E_\nu \sim 1$~GeV) might
be able to establish $CP$-violation at 3 standard deviations.
Except for certain small regions in parameter space where matter and
$CPV$ effects constructively interfere, a long-baseline
experiment with conventional superbeams would be unable to
unambiguously establish $CP$ violation because matter effects can confuse
the interpretation of the measurements.
\item{(iii)} With a sufficiently ambitious detector, if
$\sin^2 2\theta_{13} > {\cal O}(0.01)$
there is a significant region of parameter space over
which a long baseline experiment with a multi-GeV neutrino superbeam
could unambiguously establish the sign of $\delta m^2_{32}$.
\item{(iv)} Lower-energy superbeams do best at shorter distances, with a
fair reach for $\nu_\mu \to \nu_e$ appearance and some $CPV$ capability,
but little or no sensitivity to the sign of $\delta m^2_{32}$;
higher-energy superbeams do best at longer distances, with good reach
for $\nu_\mu \to \nu_e$ appearance and sign($\delta m^2_{32}$)
determination, but little or no sensitivity to $CPV$.
\item{(v)} A neutrino factory can deliver between one and two orders of
magnitude better reach in $\sin^22\theta_{13}$ for $\nu_e \to \nu_\mu$
appearance, the sign of $\delta m^2_{32}$, and $CP$ violation; for $L
\sim 3000$~km there is excellent sensitivity to all three observables.
\end{description}

Note that in this study we have restricted our considerations
to 1~GeV and multi-GeV neutrino beams. The potential of sub-GeV beams
is currently under consideration~\cite{cern guys,us}.

\section*{Acknowledgments}

This research was supported in part by the U.S.~Department of Energy
under Grants No.~DE-FG02-94ER40817, No.~DE-FG02-95ER40896 and
No.~DE-AC02-76CH03000, and in part by the University of Wisconsin Research
Committee with funds granted by the Wisconsin Alumni Research
Foundation.

\clearpage

\section*{Appendix}

To implement the Poisson statistical uncertainties in our analysis of the
$\sin^22\theta_{13}$ reach we use an approximate expression for the
upper limit ($\lambda_U$) on the number of events from the observation
of $N$ events,
\begin{equation}
        \lambda_U \simeq N  +  S \sqrt{N+1}  +  (S^2 + 2)/3  \;,
\label{eq:upper}
\end{equation}
where $S$ is the number of standard deviations corresponding to the limit.
This expression gives the correct $\lambda_U$ with an accuracy
that is better than 10\% for $N < 4$, and better than 1\% for larger
$N$~\cite{gehrels}.
If the number of predicted background events
is $B$, the expected number of signal
events corresponding to an observation 3 statistical standard deviations
above
the background is given by
\begin{equation}
         N_s  =  3 \sqrt{B+1}  +  11/3 \;.
\end{equation}
Let the systematic uncertainty on $B$ be given by
$U$.  To account
for this systematic uncertainty, we add it in quadrature with the statistical
uncertainty. Defining the quantity
\begin{equation}
        N'_s  =   \sqrt{ N_s^2  +  9U^2} \; ,
\end{equation}
the $\sin^2 2\theta_{13}$ reach can then be estimated by finding the
value of $\sin^22\theta_{13}$ that yields $N'_s$ signal events.

To determine the sign of $\delta m^2_{32}$ and/or search for $CP$
violation with conventional $\nu_\mu$ and $\bar\nu_\mu$ beams,
we will need to compare the $\nu_\mu \rightarrow \nu_e$ and
$\bar{\nu}_\mu \rightarrow \bar{\nu}_e$ appearance rates
(for a neutrino factory with a detector that measures muons, we compare
the $\nu_e \rightarrow \nu_\mu$ and $\bar{\nu}_e \rightarrow
\bar{\nu}_\mu$ appearance rates). As in the case of the
$\sin^22\theta_{13}$ reach, we will be considering the $3\sigma$
allowed regions.

Let $N$ and $\bar{N}$ be the number of events that satisfy the
signal selection criteria and are recorded
respectively during neutrino and antineutrino running.
If $N^{th}$ and $\bar N^{th}$ are theoretical predictions for
$N$ and $\bar{N}$, the region of the
$N^{th}$--$\bar N^{th}$ space allowed by the measurements is described by
\begin{equation}
\left({N^{th} - N\over\alpha_N}\right)^2 +
\left({\bar N^{th} - \bar N\over\alpha_{\bar N}}\right)^2 \le 1 \,,
\label{eq:a1}
\end{equation}
where $\alpha_N$ and $\alpha_{\bar N}$ are the experimental
uncertainties on $N$ and $\bar N$, respectively.
In the absence of systematic uncertainties, and in the approximation of
Gaussian statistics, the $3\sigma$ values
are $\alpha_N = 3\sqrt{N}$ and $\alpha_{\bar N} = 3\sqrt{\bar N}$.
However, since $N$ and $\bar{N}$ might be small Gaussian statistics
may be inappropriate. Instead,
we define $\alpha_N$ and $\alpha_{\bar{N}}$ to
correspond to the appropriate 99.87\% confidence level deviations from
the central values of $N$ and $\bar{N}$, respectively, using
Poisson statistics. The expressions for $\alpha_{N}$ and
$\alpha_{\bar{N}}$ will depend on whether we are considering an
upper or lower limit.

Consider first the case of an upper limit.  We can compute $\alpha_{N}$
using Eq.~(\ref{eq:upper}), with $S = 3$, yielding:
\begin{equation}
\alpha_{N}^{upper} = 3 \sqrt{N + 1} + 11/3 \,.
\label{eq:a2}
\end{equation}
To compute the value for $\alpha_{\bar{N}}$ for a lower bound,
we need an expression for the Poisson lower limit given the observation
of $N$ events. We use the expression from Ref.~\cite{gehrels}, namely:
\begin{equation}
\lambda_L \; \simeq \; N \left(
1 - \frac{1}{9N} - \frac{S}{3\sqrt{N}} + \beta N^\gamma
\right)^3 \; ,
\label{eq:a3}
\end{equation}
where with $S=3$ we have $\beta = 0.222$ and $\gamma = -1.88$.
This approximate expression for the Poisson lower limit on $N$
is accurate to a few percent or better for all $N$.
Hence
\begin{equation}
\alpha_N^{lower} = N \; - \; N
\left( 1 - \frac{1}{9N} - \frac{3}{3\sqrt{N}} +
0.222 N^{-1.88}
\right)^3 \; .
\label{eq:a4}
\end{equation}
The corresponding values for $\alpha_{\bar N}$ can be found by
substituting $\bar N$ for $N$ in Eqs.~(\ref{eq:a2}) and (\ref{eq:a4}).

In practice $N$ and $\bar{N}$ will contain
background components $B$ and $\bar{B}$.
The predicted backgrounds will have associated systematic uncertainties
$U$ and $\bar{U}$.
In this case we can still use Eq.~(\ref{eq:a1}) to determine the allowed
regions, but to take account of the background and systematic uncertainties
the $\alpha^2_N$ and $\alpha^2_{\bar{N}}$
are replaced with the substitutions:
\begin{equation}
\alpha^2_{N} \; \rightarrow \; \alpha^2_{N} + 9U^2
\label{eqn:a5}
\end{equation}
\begin{equation}
\alpha^2_{\bar{N}} \; \rightarrow \; \alpha^2_{\bar N} +
9\bar{U}^2 \; .
\label{eqn:a6}
\end{equation}
Other systematic uncertainties on the predicted $N$ and $\bar{N}$
(for example, the uncertainty on the neutrino and antineutrino
cross-sections) can be handled in a similar way, by
replacing $\alpha_N$ ($\alpha_{\bar{N}}$) with the
quadrature sum of $\alpha_N$ ($\alpha_{\bar{N}}$) and the
additional 99.87\% C.L. uncertainty on $N$ ($\bar{N}$).

\clearpage

\begin{table}[t]
\caption[]{\label{tab:xx}
Parameters for scenarios $A$, $F$, and $W$ discussed in the text.
           The scenarios assume 3 years of neutrino running and
           6 years of antineutrino running. The dataset sizes $D$
and background fractions $f_B$ are defined for the event samples after
the signal selection requirements have been applied.}
\begin{tabular}{lcccccc}
    &  \multicolumn{2}{c}{Scenario $A$}
    & \multicolumn{2}{c}{Scenario $F$}
    & \multicolumn{2}{c}{Scenario $W$}\\
    & $\nu$& $\bar\nu$& $\nu$& $\bar\nu$& $\nu$& $\bar\nu$ \\
\hline
%Beam flux&  $4 \times\rm NUMI$& $4 \times\rm NUMI$& $4 \times\rm NUMI$&
%$4 \times\rm NUMI$\\
Fiducial mass (kt)&   30&	30& 10 & 10 &      220&	  220\\
$D$ (kt-years)&	      45&	90& 27 & 54 &      450&	  900\\
Backg. frac. $f_B$& 0.004& 0.006& 0.013 & 0.015 &     0.02&	  0.02\\
Backg. uncertainty $\sigma_{f_B}/f_B$&   0.1&	0.1& 0.1& 0.1& 0.1& 0.1
\end{tabular}
\end{table}

\begin{table}
\caption[]{\label{tab:x}
The fraction of pions decaying in a channel with the maximum
length permitted by the depth of viable rock ($D_{max}$) under the
accelerator site, tabulated as a function of baseline $L$ for
configurations with and without a near detector.}
\begin{tabular}{lccccc}
SNuMI& $E_\nu$(peak)& $D_{\rm max}$& $L$& \multicolumn{2}{c}{$f_{\rm decay}$}\\
Beam &      (GeV)&         (m)&      (km)&    with near&      no near\\
\hline
LE&         3&	          200&	    2900&       0.93&	        0.95\\
  &          &               &      7300&       0.36&           0.56\\
  &          &               &      9300&        ---&           0.37\\
\hline
ME&         7&            200&	    2900&       0.67&	        0.72\\
  &          &               &      7300&       0.17&           0.29\\
  &          &               &      9300&        ---&           0.17\\
\hline
HE&        15&            200&      2900&       0.41&           0.44\\
  &          &               &      7300&       0.09&           0.16\\
  &          &               &      9300&        ---&           0.09\\
\hline
\hline
LE&         3&            400&	    2900&       0.98&           0.99\\
  &          &               &      7300&       0.82&           0.88\\
  &          &               &      9300&       0.67&           0.77\\
\hline
ME&         7&	          400&	    2900&       0.93&           0.94\\
  &          &               &      7300&       0.56&           0.64\\
  &          &               &      9300&       0.38&           0.47\\
\hline
HE&        15&	          400&	    2900&       0.69&           0.72\\
  &          &               &      7300&       0.29&           0.34\\
  &          &               &      9300&       0.20&           0.26
\end{tabular}
\end{table}

\begin{table}[t]
\caption[]{\label{tab:y}
$\sin^2 2\theta_{13}$ reach (corresponding to a signal that is 3 standard
deviations above the background after 3 years of running with a
neutrino superbeam) shown as a function of baseline $L$
for Scenario $A$ described in the text. The oscillation probability
$P(\nu_\mu \rightarrow \nu_e)$ corresponding to $\sin^2 2\theta_{13} = 0.01$
and the expected numbers of signal
events $S$ and background events $B$ are also listed.
The calculations assume $\Delta m^2_{32} = 3.5 \times 10^{-3}$~eV$^2$,
$\delta m^2_{21} = 5 \times 10^{-5}$~eV$^2$, and $\delta = 0$.}
\begin{tabular}{lcrcccc}
SNuMI&  $E_\nu$(peak)&  $L$&  $P$&  $S$&   $B$&  $\sin^2 2\theta_{13}$ reach\\
Beam &      (GeV)&       (km)& \\
\hline
LE&          3&	        730&   0.0024&  210&    340&     0.006\\
  &           &        2900&   0.0045&   26&     24&     0.008\\
  &           &        7300&    0.012&  4.1&    1.3&     0.02\\
  &           & 9300\rlap{(*)}& 0.016&  3.2&    0.8&     0.02\\
\hline
ME&          7&	        730&    0.0016&	 370&   910&    0.01\\
  &           &        2900&    0.0075&  120&    62&	 0.003\\
  &           &        7300&     0.025&	 15&     2.4&    0.006\\
  &           &  9300\rlap{(*)}& 0.035&	 14&     1.6&    0.006\\
\hline
HE&         15&	        730&    0.0006&	 290&   2000&    0.02\\
  &           &        2900&    0.0054&	 180&    130&    0.003\\
  &           &        7300&     0.024&	 25&     4.2&    0.004\\
  &           &  9300\rlap{(*)}& 0.032&	 25&     3.1&	 0.004
\end{tabular}
(*)  No near detector
\end{table}

\begin{table}[t]
\caption[]{\label{tab:z}
$\sin^2 2\theta_{13}$ reach (corresponding to a signal that is 3 standard
deviations above the background after 3 years of running with
a neutrino superbeam) shown as a function of baseline $L$
for Scenario $W$  described in the text. The oscillation probability
$P(\nu_\mu \rightarrow \nu_e)$ corresponding to $\sin^2 2\theta_{13} = 0.01$,
and the expected numbers of signal
events $S$ and background events $B$ are also listed.
The calculations assume $\Delta m^2_{32} = 3.5 \times 10^{-3}$~eV$^2$,
$\delta m^2_{21} = 5 \times 10^{-5}$~eV$^2$, and $\delta = 0$.}
\begin{tabular}{lcrcccc}
SNuMI&  $E_\nu$(peak)&  $L$&  $P$&  $S$&  $B$& $\sin^2 2\theta_{13}$ reach\\
Beam &      (GeV)&      (km)& \\
\hline
LE&           3&        730&  0.0024&  2100&  17000&     0.03\\
  &            &       2900&  0.0045&   260&   1200&     0.02\\
  &            &       7300&   0.012&    41&    67&     0.009\\
  &            & 9300\rlap{(*)}& 0.016&	 32&    40&     0.008\\
\hline
ME&           7&        730&   0.0016&	3700&	46000&	0.05\\
  &            &       2900&   0.0075&	1200&	3100&	0.008\\
  &            &       7300&    0.024&	150&	120&	0.003\\
  &            & 9300\rlap{(*)}& 0.035&	140&	80&	0.003\\
\hline
HE&          15&       730&     0.0006&	2900&	98000&	0.1\\
  &            &      2900&     0.0054&	1800&	6700&	0.01\\
  &            &      7300&	0.024&	250&	210&	0.003\\
  &            & 9300\rlap{(*)}& 0.032&	250&	160&	0.003
\end{tabular}
(*)  No near detector
\end{table}

\begin{table}[t]
\caption[]{\label{tab:s}
Summary of the $\sin^2 2\theta_{13}$ reach (in units of $10^{-3}$)
for various combinations of neutrino beam,
distance, and detector for
(i) a $3\sigma$ $\nu_\mu \to \nu_e$ appearance with $\delta m^2_{21}
= 10^{-5}$~eV$^2$,
(ii) a $3\sigma$ determination of the sign of $\delta m^2_{32}$ with
$\delta m^2_{21} = 5\times10^{-5}$~eV$^2$ , and
(iii) a $3\sigma$ discovery of $CP$ violation for three values of
$\delta m^2_{21}$ (in eV$^2$). Dashes in the sign of $\delta m^2_{32}$ column
indicate that the sign is not always determinable. Dashes in the $CPV$
columns indicate $CPV$ cannot be established for $\sin^22\theta_{13}
\le 0.1$, the current experimental upper limit, for any values of the
other parameters. The $CPV$ entries are calculated assuming the value of
$\delta$ that gives the maximal disparity of $N(e^+)$ and $N(e^-)$; for
other values of $\delta$, $CP$ violation may not be measurable.}
\begin{tabular}{ccc|c|c|ccc}
&&& \multicolumn{5}{c}{$\sin^2 2\theta_{13}$ reach (in units of $10^{-3}$)}\\
\cline{4-8}
&&& $\nu_\mu \to \nu_e$ & Unambiguous &
\multicolumn{3}{c}{Possible $3\sigma$ $CPV$ }\\
&&& appearance & $3\sigma$ sign($\delta m^2_{32}$) &
\multicolumn{3}{c}{$\delta m^2_{21}$ (in eV$^2$)}\\
 Beam&  $L$ (km) &  Detector & $\delta m^2_{21} = 10^{-5}$
& $\delta m^2_{21} = 5\times10^{-5}$ & $5\times10^{-5}$
& $1\times10^{-4}$ & $2\times10^{-4}$\\
\hline\hline
JHF  & 295 & A & $25$ & $-$ & $-$ &  $-$ & $25$\\
     &     & W & $17$ & $-$ & $-$ & $40$ & $8$\\
\hline\hline
SJHF & 295 & A &  $8$ & $-$ &   $-$ &  $5$ & $3$\\
     &     & W & $15$ & $-$ & $100$ & $20$ & $5$\\
\hline\hline
SNuMI LE & 730 & A &  $7$ & $-$ & $100$ & $20$ &  $4$\\
         &     & W & $30$ & $-$ &   $-$ &  $-$ & $40$\\
\hline\hline
SNuMI ME & 2900 & A & $3$ & $6$ & $-$ & $-$ & $100$\\
         &      & W & $8$ & $15$ & $-$ & $-$ &   $-$\\
\hline
         & 7300 & A &  $6$ & $6$ & $-$ & $-$ & $-$\\
         &      & W &  $3$ & $3$ & $-$ & $-$ & $-$\\
\hline\hline
SNuMI HE & 2900 & A &  $3$ & $7$ & $-$ & $100$ & $20$\\
         &      & W & $10$ & $15$ & $-$ &   $-$ &   $-$\\
\hline
         & 7300 & A &  $4$ & $4$ & $-$ & $-$ & $-$\\
         &      & W &  $3$ & $3$ & $-$ & $-$ & $-$\\
\hline\hline
20 GeV NuF                 & 2900 & 50~kt & $0.5$ & $2.5$ & $-$ & $2$ & $1.5$\\
$1.8\times10^{20}$ $\mu^+$ & 7300 &       & $0.5$ & $0.3$ & $-$ & $-$ & $-$\\
\hline
20 GeV NuF             & 2900 & 50~kt & $0.1$ & $1.2$ & $0.6$ & $0.4$ & $0.6$\\
$1.8\times10^{21}$ $\mu^+$ & 7300 &   & $0.07$ & $0.1$ & $-$  & $-$   & $-$\\
\end{tabular}
\end{table}

\clearpage

% 1

% 2

% 3

% 4

% 5

%  new 6

% 7

% 8

% 9

% 10

% 11

% 12

\end{document}